\documentclass[twocolumn,showpacs,preprintnumbers,amsmath,amssymb,APSl,prd,nofootinbib,superscriptaddress]{revtex4-1}

\usepackage{dcolumn}
\usepackage{bm}
\usepackage{ifpdf}
\usepackage{hyperref}
\usepackage{bm}
\usepackage{xcolor,color,graphicx,graphics}
\usepackage[spanish,english]{babel}
\usepackage[latin1]{inputenc}
\usepackage[OT1]{fontenc}
\usepackage{latexsym,amssymb,amsmath,amsfonts}
\usepackage{makeidx}
\usepackage{epsfig,subfigure}
\usepackage{natbib}
\usepackage{epstopdf}
\usepackage{mathrsfs}
\usepackage{hyperref}
\hypersetup{colorlinks=true, linkcolor=blue, citecolor=green}
\usepackage{enumerate}

\definecolor{red}{rgb}{1,0,0}

\def\+{^\dagger}

\def\<{\leftarrow}
\def\>{\rightarrow}

\def\({\left(}
\def\){\right)}

\newcommand{\bi}{\begin{itemize}} 				\newcommand{\ei}{\end{itemize}}
\newcommand{\benu}{\begin{enumerate}} 		\newcommand{\enu}{\end{enumerate}}
\newcommand{\bd}{\begin{dinglist}{0}}     \newcommand{\ed}{\end{dinglist}}
\newcommand{\bfig}{\begin{figure}[htbp]}  \newcommand{\efig}{\end{figure}}
        			
\newcommand{\bc}{\begin{center}} 				  \newcommand{\ec}{\end{center}}
\newcommand{\be}{\begin{equation}} 				\newcommand{\ee}{\end{equation}}
\newcommand{\bsub}{\begin{subequations}}  \newcommand{\esub}{\end{subequations}}
\newcommand{\ben}{\begin{eqnarray}} 			\newcommand{\een}{\end{eqnarray}}
\newcommand{\ba}[1]{\begin{array}{#1}} 		\newcommand{\ea}{\end{array}}
\newcommand{\bea}{\begin{equation}\begin{array}{rcl}}
\newcommand{\eea}{\end{array}\end{equation}}

\begin{document}

\title{Cosmological bounces, cyclic universes, and effective cosmological constant in Einstein-Cartan-Dirac-Maxwell theory}

\author{Francisco Cabral}
\email{ftcabral@fc.ul.pt}
\affiliation{Instituto de Astrof\'{\i}sica e Ci\^{e}ncias do Espa\c{c}o, Faculdade de
Ci\^encias da Universidade de Lisboa, Edif\'{\i}cio C8, Campo Grande,
P-1749-016 Lisbon, Portugal}
\author{Francisco S. N. Lobo} \email{fslobo@fc.ul.pt}
\affiliation{Instituto de Astrof\'{\i}sica e Ci\^{e}ncias do Espa\c{c}o, Faculdade de
Ci\^encias da Universidade de Lisboa, Edif\'{\i}cio C8, Campo Grande,
P-1749-016 Lisbon, Portugal}
\author{Diego Rubiera-Garcia} \email{drubiera@ucm.es}
\affiliation{Departamento de F\'isica Te\'orica and IPARCOS, Universidad Complutense de Madrid, E-28040
Madrid, Spain}

\date{\today}

\begin{abstract}
Einstein-Cartan theory is an extension of the standard formulation of General Relativity characterized by a non-vanishing torsion. The latter is sourced by the matter fields via the spin tensor, and its effects are expected to be important at very high spin densities. In this work we analyze in detail the physics of Einstein-Cartan theory with Dirac and Maxwell fields minimally coupled to the spacetime torsion. This breaks the $U(1)$ gauge symmetry, which is suggested by the possibility of  a torsion-induced phase transition in the early Universe. The resulting Dirac-like and Maxwell-like equations are non-linear with self-interactions as well as having fermion-boson non-minimal couplings. We discuss several cosmological aspects of this theory under the assumption of randomly oriented spin densities (unpolarized matter), including bounces, acceleration phases and matter-antimatter asymmetry in the torsion era, as well as late-time effects such as the generation of an effective cosmological constant, dark energy, and future bounces within cyclic solutions.
\end{abstract}


\maketitle

\section{Introduction}

The standard model of particles and interactions is an extremely successful theoretical construction, being able to describe the phenomena that we observe with current detectors in particle accelerator collisions and in cosmic rays. It rests deeply on i) (quantum) gauge field theories, which reveal a fundamental role of symmetry principles in the physics of interactions, and ii) on the rigid four-dimensional flat (Minkowski) spacetime background of special relativity, and as such it does not include gravity. The physics of particles and interactions of the early Universe is extrapolated from the success of this paradigm to describing the phenomena up to very high densities and temperatures at the electroweak scale.  In the very early Universe one should incorporate strong-field gravitational effects, which requires new ideas in order to unveil the nature of the gravitational interaction on such scales.

Einstein's theory of General Relativity (GR) rests on a fundamental geometrical principle according to which gravity is deeply connected to spacetime and, in particular, to its geometry. Built upon this principle, GR has passed many tests from Solar System to binary pulsars \cite{Will:2014kxa}, stellar orbits around the central galactic black hole \cite{SgrA2017PhRvL.118u1101H}, gravitational waves (GWs) from coalescing compact objects \cite{Abbott:2016blz,Abbott:2017nn,TheLIGOScientific:2016src,Abbott:2017oio,TheLIGOScientific:2017qsa}, the indirect observation of the black hole's horizon (through its effects on surrounding radiation and plasma) with the Event Horizon Telescope \cite{M87EventHorizon}, and to some extent the cosmological observations \cite{Bull:2015stt}. Indeed, the topics of dark matter and dark energy as well as the initial conditions and big bang singularity are still fundamental open questions that drive extensive research efforts. In many cases, such efforts involve extensions of GR, which have been confronted against the first results from GW astronomy \cite{Ezquiaga:2017ekz}. Several relevant reviews of these theories and related phenomenology can be found, for instance, in Refs. \cite{DeFelice:2010aj,Olmoreview, CLreview,NOOreview,Berti:2015itd,BeltranJimenez:2017doy,Harko:2014gwa,Capozziello:2015lza,BookHarkoLobo}.

Both the amazing successes of symmetry principles in the physics of interactions and the geometrical methods in gravity can be consistently combined by extending the gauge principle to gravity. Gauge theories of gravity reveal indeed a deep connection between spacetime symmetries and spacetime (non-Euclidean) geometries. Relevant groups of spacetime transformations (changes in spacetime coordinates) such as the Poincar\'{e} P(1,3), the Weyl W(1,3), the conformal C(1,3), the general linear GL(4,$\Re$), and the affine A(4,$\Re$) groups lead to different spacetime geometries and of theories of gravity. This is done by imposing the local symmetry of the matter Lagrangian upon the acting of such transformations and by constructing the gravitational Lagrangian with the (invariant) field strengths corresponding to the gravitational gauge potentials. These field strengths turn out to be well known geometrical objects such as curvature, torsion and nonmetricity. Theories of gravity with richer spacetime geometries can predict new physics beyond GR in strong gravity regimes and also at the linearized weak-field limits. For detailed reviews on gauge theories of gravity and their applications see \cite{Blagojevic:2013xpa,Blagojevic,Obukhov}.

In the early Universe, the standard model of particles and interactions leads to the ideas of symmetry breaking phase transitions and Higgs-like mechanisms depending on some critical parameters, such as the temperature. Similarly, different symmetries in the gauge theories of gravity can be unified into higher symmetry groups or broken into smaller ones. For instance, the conformal group includes the Poincar\'{e} group together with scale transformations (Weyl rescaling) and proper conformal transformations, and can be broken into the Poincar\'{e} group. In these symmetry breakings, the corresponding phase transitions affect both the gravitational and matter field degrees of freedom \cite{Linde:1978px}. Indeed, phase transitions in the early universe, induced by these theories of gravity, are expected to leave imprints, for instance, in the particle physics of the quark-gluon-lepton plasma \cite{Zakout:2006zj}. These imprints could be probed by cosmological GWs \cite{Grojean:2006bp}, neutrino \cite{Lello:2014yha} and radiation (CMB) \cite{Barriga:2000nk} backgrounds. This may have a profound impact on scale-invariance regimes and its symmetry breaking, parity breaking \cite{Blagojevic:2013xpa,Blagojevic,Obukhov}, CP breaking and matter/anti-matter asymmetries \cite{Poplawski:2011xf,Obukhov:2018bmf}, $U(1)$-gauge breaking \cite{Cabral:2019gzh}, Higgs-like mechanisms, etc.

It is reasonable to assume that classical gauge theories of gravity such as the metric-affine theories or Poincar\'{e} gauge theories of gravity (PGTG \cite{Obukhov:2006gea,Obukhov:2018bmf}) are effective, low-energy limits of a more fundamental quantum gravity theory. The PGTG class is fundamental (given the importance of the Poincar\'{e} symmetries in relativistic field theories) and the most general quadratic Lagrangian ({\it $\grave{a}$ la} Yang-Mills) contains parity breaking terms induced by the richer Riemann-Cartan (RC) geometry with curvature and torsion \cite{Obukhov:2018bmf}. The simplest PGTG is  Einstein-Cartan-Sciama-Kibble (ECSK) theory \cite{Blagojevic:2013xpa,Blagojevic,Obukhov}, which predicts torsion effects at very high energy densities via an algebraic relation between the spin density of matter fields and spacetime torsion. The latter affects the Einstein-like equations for the metric but also the dynamics of fermions and bosons coupled to gravity.
Although the effects upon the metric are expected to be relevant only at extreme densities, such as those found in the early universe or inside black holes, the effects on the matter fields can be important in the deep interior of compact objects such as magnetars or hypothetical quark stars. In cosmology, these effects are relevant for the physics around the Grand Unification phase transition scale and beyond, and one speaks of a torsion era, where the corresponding energy density is expected to scale with $\sim a^{-6}$. Indeed, theories with torsion in cosmological scenarios have been thoroughly studied in the literature for decades, with a large pool of applications \cite{Poplawski:2011jz,Unger:2018oqo,Kranas:2018jdc,Poplawski:2010kb,Ivanov:2016xjm,Razina:2010bj,Palle:2014goa, Poplawski:2012qy,Xue:2008qs,Vakili:2013fra}, as well as for their $f(T)$ extensions \cite{Cai:2015emx,Harko:2014sja,Chen:2010va,Rodrigues:2012qua,Li:2011wu,Bamba:2012vg,Harko:2014sja,Harko:2014aja,Carloni:2015lsa,Saez-Gomez:2016wxb}.

In a previous work \cite{Cabral:2019gzh} we considered an extension of the ECSK theory by adding a minimal coupling to fermionic (Dirac) and bosonic (Maxwell) fields in the RC geometry. The resulting Einstein-Cartan-Dirac-Maxwell (ECDM) model contains new non-linear generalized Dirac-Hehl-Data and electromagnetic equations with non-minimal interactions between fermionic and bosonic fields. While the coupling to Dirac fields has been considered previously in the literature, for instance within particle physics \cite{Khanapurkar:2018jvx,Khanapurkar:2018gyo,Lucat:2015rla,Poplawski:2011wj,Poplawski:2010jv,Poplawski:2009su}, the minimal coupling of Maxwell fields to torsion breaks the $U(1)$ gauge symmetry and, therefore, it is usually assumed that torsion simply does not couple to electromagnetic fields. Any new physics in extreme environments, linked to a $U(1)$ symmetry breaking phase transition induced by torsion is therefore unexplored and, as far as we know, there are no observational constraints in the literature, though we point out that ECSK leaves the vacuum dynamics unaffected, implying its consistency with solar system experiments. In general, cosmological torsion effects are expected to dominate when Cartan's density, $\rho_C \sim 10^{54}$g/cm$^3$ is approached  and, therefore, the ECDM model allows to implement a valid physical mechanism to generate cosmological phase transitions during the torsion era \cite{Cabral:2019gzh}. 

The main aim of the present paper is therefore to derive the cosmological equations governing the geometry and the matter fields within ECDM theory, and to study thoroughly their consequences for the dynamics of the universe. The latter involve the existence of cosmological bounces and the generation of an effective cosmological constant, which were already present in the usual ECSK theory, as well as some novelties under the form of acceleration/desacceleration phases, matter-antimatter asymmetry, and late-time effects including the existence of cyclic cosmologies with a desacceleration period ended in a bounce and followed by a new acceleration period. These findings highlight the richness of the theories with torsion in yielding new features of interest not present in the $\Lambda$CDM concordance model.

This paper is organized as follows.
In Sec. \ref{secII}, mostly supported on the results of Ref.\cite{Cabral:2019gzh}, we introduce ECDM gravity, paying special attention to the minimal couplings between torsion and matter represented by classical fundamental bosonic and fermionic fields and deriving the gravitational field equations. Sec. \ref{secIV} contains the core results of this work, including the modified Friedman equations and related phenomenology of interest within different cosmological regimes, as well as the dynamics of bosonic (Maxwell) and fermionic (Dirac) fields in the cosmological framework. Finally, in Sec. \ref{secV}, we summarize our findings and further discuss and interpret our results.

\section{Einstein-Cartan-Dirac-Maxwell theory}\label{secII}

\subsection{Einstein-Cartan-Sciama-Kibble theory}

The ECSK theory and its ECDM extension considered in this work are endowed with a Riemann-Cartan (RC) geometry (see \cite{Blagojevic:2013xpa,Obukhov}), i.e., with curvature and torsion $T^{\alpha}{}_{\mu\nu}\equiv\Gamma^{\alpha}{}_{[\mu\nu]}$, and its action can be written as
\begin{equation} \label{eq:actionEC}
S_{\rm ECSK}=\dfrac{1}{2\kappa^2} \int d^4x \sqrt{-g}R + \int d^4x \sqrt{-g}\,\mathcal{L}_{m} \ ,
\end{equation}
with the following definitions and conventions: $\kappa^2=8\pi G$, where $G$ is Newton's constant, $g$ is the determinant of the spacetime metric $g_{\mu\nu}$, the curvature scalar $R=g_{\mu\nu}R^{\mu\nu}$ is constructed out of the Ricci tensor $R_{\mu\nu}(\Gamma) \equiv R^{\alpha}{}_{\mu\alpha\nu}(\Gamma)$, and the matter Lagrangian $\mathcal{L}_m=\mathcal{L}_{m}(g_{\mu\nu},\Gamma,\psi_m)$ depends on the metric and the matter fields $\psi_m$  and also on contortion $K_{\alpha\mu\nu} \equiv T_{\alpha\mu\nu}+2T_{(\mu\nu)\alpha}$ via the covariant derivatives of the matter fields. We note that in the above action the fact that the RC connection, $\Gamma_{\mu\nu}^{\lambda}=\tilde{\Gamma}_{\mu\nu}^{\lambda}+K_{\mu\nu}^{\lambda}$\footnote{Unless stated otherwise, in this work all quantities with a tilde represent the corresponding expressions computed in GR (curved, pseudo-Riemann spacetime). }, has an antisymmetric part, yields new contributions to the standard Einstein equations and to the dynamical equations for the matter fields.

From the perspective of a gauge formulation of gravity and the metric-affine approach, the theory is consistently formulated with the tetrads and spin (Lorentz) connection variables ($\vartheta^{a},w_{ab}$) representing the gravity or geometric degrees of freedom which are associated to the usual spacetime metric and affine connection variables. The field equations are obtained by independent variation of the above action with respect to metric and contortion (or, equivalently, with respect to the tetrads and the Lorentz spin connection) and the matter fields. The corresponding Einstein-like equations can then be cast as
\begin{equation} \label{eq:EOMeff}
\tilde{G}_{\mu\nu}=\kappa^{2}T^{\rm eff}_{\mu\nu} \ ,
\end{equation}
where $\tilde{G}_{\mu\nu}$ is the usual Einstein tensor, and $T^{\rm eff}_{\mu\nu}=T_{\mu\nu}+U_{\mu\nu}$ is the effective energy-momentum tensor composed of the usual dynamical energy-momentum piece
$T_{\mu\nu}$, while the $U_{\mu\nu}$ term emerges from the quadratic torsion corrections to the Ricci tensor and introduces corrections quadratic in the contortion (or torsion). It is important to note that in ECSK theory torsion does not propagate and obeys the Cartan equations
\begin{equation} \label{eq:cartan}
T^{\alpha}_{\;\beta\gamma}+\delta^{\alpha}_{\beta}T_{\gamma}-\delta^{\alpha}_{\gamma}T_{\beta}=\kappa^2 s^{\alpha}_{\;\beta\gamma} \ ,
\end{equation}
or
\begin{equation}
T^{\alpha}_{\;\beta\gamma}=\kappa^2\left( s^{\alpha}_{\;\beta\gamma} +\delta^{\alpha}_{[\beta}s_{\gamma]}\right) ,
\end{equation}
where $s^{\alpha\mu\nu}\equiv\dfrac{\delta L_{m}}{\delta K_{\mu\nu\alpha}}$ is the spin tensor of matter, with dimensions of energy/area or spin/volume. Since in absence of matter the spin density tensor vanishes, this implies that ECDM theories does not propagate additional degrees of freedom beyond the two polarizations of the gravitational field of GR and, as consequence, it is ghost-free and not affected by any of the associated instabilities (for a detailed analysis of this question for general metric-affine theories see \cite{Jimenez:2020dpn}).

\subsection{Matter Lagrangian and torsion from fundamental fermionic and bosonic fields}

In this subsection we shall briefly summarize the main steps to derive the matter field dynamics and discuss the most salient features at the Lagrangian level, in order to pave the ground for the new results obtained in this work. For a more detailed derivation see Ref. \cite{Cabral:2019gzh}. Thus we consider a minimal coupling between torsion and matter, the latter represented here by classical bosonic (four-vector) and fermionic (four-spinor) fields. This can be directly implemented at the level of the matter Lagrangian as
\begin{equation}
\label{MatterLag}
\mathcal{L}_{m}=\mathcal{L}_{\rm D}+\mathcal{L}_{\rm M}+j^{\mu}A_{\mu} \ ,
\end{equation}
where  $j^{\mu}=q\bar{\psi}\gamma^{\mu}\psi$ is Dirac's four-current and
\begin{equation}
\label{DiracLag}
\mathcal{L}_{\rm D}=\tilde{\mathcal{L}}_{\rm D}+
K_{\alpha\beta\mu}s^{\mu\alpha\beta}_{D} \ ,
\end{equation}
is the Dirac Lagrangian with minimal coupling to the geometry of RC spacetime\footnote{This expression is derived from the Dirac Lagrangian density
\begin{equation}
\mathcal{L}_{\rm Dirac}=\dfrac{i\hbar}{2}\left(\bar{\psi}\gamma^{\mu}D_{\mu}\psi-(D_{\mu}\bar{\psi})\gamma^{\mu}\psi\right)-m\bar{\psi}\psi \ , \nonumber
\end{equation}
for spinors $\psi$ and their adjoints $\bar{\psi}=\psi^{+}\gamma^0$, where the covariant derivatives are defined as
\begin{eqnarray}
D_{\mu}\psi&=&\tilde{D}_{\mu}\psi+\dfrac{1}{4}K_{\alpha\beta\mu}\gamma^{\alpha}\gamma^{\beta}\psi \ , \nonumber \\
D_{\mu}\bar{\psi}&=&\tilde{D}_{\mu}\bar{\psi}-\dfrac{1}{4}K_{\alpha\beta\mu}\bar{\psi}\gamma^{\alpha}\gamma^{\beta} \ , \nonumber
\end{eqnarray}
with $D_{\mu}$ and $\tilde{D}_{\mu}$ being the (Fock-Ivanenko) covariant derivatives built with the Cartan connection and the Levi-Civita connection, respectively, and $\gamma^{\mu}$ are the induced Dirac-Pauli matrices which obey $\left\lbrace \gamma^{\mu},\gamma^{\nu}\right\rbrace =2g^{\mu\nu}I$, where $I$ is the $4\times4$ unit matrix. Taking $\tilde{\mathcal{L}}_{\rm D}$ we get the expression for Dirac's Lagrangian in curved spacetime. }, and $s_{\;\;\alpha\beta\gamma}^{D}=\dfrac{1}{2}\epsilon_{\alpha\beta\gamma\lambda}\breve{s}^{\lambda}$ is the totally antisymmetric Dirac spin tensor, expressed in terms of the axial (spin) vector $\breve{s}^{\lambda}=\dfrac{\hbar}{2}\bar{\psi}\gamma^{\lambda}\gamma^{5}\psi$.

Similarly, bosons are represented by the generalized Maxwell Lagrangian in a RC spacetime written as
\begin{equation}\label{eq:Maxfull}
\mathcal{L}_{\rm M}=\frac{\lambda}{4}F_{\mu\nu}F^{\mu\nu} \,,
\end{equation}
where $\lambda$ is a coupling parameter setting the system of units. The generalized field strength tensor is defined as
\begin{equation}
\label{newFaraday}
F_{\mu\nu}\equiv \nabla_{\mu}A_{\nu}-\nabla_{\nu}A_{\mu}=\tilde{F}_{\mu\nu}+2K^{\lambda}_{\;\;[\mu\nu]}A_{\lambda} \ , \nonumber
\end{equation}
where $\nabla_\mu$ is the covariant derivative in RC spacetime constructed with the Cartan connection, while $\tilde{F}_{\mu\nu}=\partial_{\mu}A_{\nu}-\partial_{\nu}A_{\mu}$ is the standard field strength tensor when torsion is neglected.
%
More explicitly, Eq.(\ref{eq:Maxfull}) can be expressed as
\begin{equation}
\label{newMaxLagra}
\mathcal{L}_{\rm M}=\tilde{\mathcal{L}}_{\rm M}+\lambda \left(T^{\lambda\mu\nu}T^{\gamma}_{\;\;\mu\nu}A_{\gamma}+T^{\lambda\mu\nu}\tilde{F}_{\mu\nu} \right)A_{\lambda} \ .
\end{equation}
We see that the  torsion contribution explicitly breaks the $U(1)$ symmetry of (Maxwell) massless four-vector fields coupled to (Dirac) fermions.
For this matter Lagrangian, the Cartan equations (\ref{eq:cartan}) become
\begin{equation}
\label{torsionextended}
T^{\alpha}{}_{\beta\gamma}=\kappa^2\Big( s_{\;\;\;\;\beta\gamma}^{D\;\alpha}+ s_{\;\;\;\;\beta\gamma}^{M\;\alpha}+\delta^{\alpha}_{[\beta}s^{M}_{\gamma]}\Big) \ ,
\end{equation}
where we have broken down the total spin
$s_{\lambda\alpha\beta}=s^{M}_{\lambda\alpha\beta}+s^{D}_{\lambda\alpha\beta}$, in terms of its electromagnetic
\begin{eqnarray}
s_{\lambda\mu\nu}^{M}&\equiv&\frac{\delta \mathcal{L}^{\rm corr}_{\rm M}}{\delta K^{\mu\nu\lambda}}=\lambda A_{[\mu}F_{\nu]\lambda}
\nonumber \\
&=&\lambda\left(A_{[\mu}\tilde{F}_{\nu]\lambda}+2A_{[\mu}T^{\alpha}{}_{\;\nu]\lambda}A_{\alpha} \right)\,,
\end{eqnarray}
and Dirac contributions (to be explicitly computed on each case). After some algebra, Cartan's equations (\ref{torsionextended}) yield torsion as a function of the fermionic and bosonic fields as
\begin{eqnarray}
\label{finalnewCartan}
T^{\alpha}{}_{\beta\gamma}
&=&\kappa^{2}\Big[\tilde{s}^{M\alpha}{}_{\beta\gamma}+
s^{D\alpha}{}_{\beta\gamma}+2\lambda\kappa^{2}s^{D\alpha\;\;\;\;\rho}_{\;\;\;[\beta}A_{\gamma]}A_{\rho}
\nonumber \\
&&+\dfrac{2}{2+\lambda\kappa^{2}A^{2}}\left(\delta^{\alpha}_{[\beta}\tilde{s}^{M}_{\gamma]}-\lambda\kappa^{2}A^{\alpha}A_{[\beta}\tilde{s}^{M}_{\gamma]}\right)\Big] \ ,
\end{eqnarray}
with $\tilde{s}_{\lambda\alpha\beta}\equiv\lambda A_{[\alpha}\tilde{F}_{\beta]\lambda}$ and $\tilde{s}_{\nu}=\tilde{s}^{\alpha}_{\;\;\nu\alpha}$.
A particular case of this expression is the one of  fermionic torsion
\begin{equation}
\label{fermionictorsion}
T^{\alpha}{}_{\beta\gamma}
=\kappa^{2}s^{D\alpha}{}_{\beta\gamma}, \qquad s_{\;\;\alpha\beta\gamma}^{D}=\dfrac{1}{2}\epsilon_{\alpha\beta\gamma\lambda}\breve{s}^{\lambda} \ ,
\end{equation}
that is, torsion being exclusively the result of fermionic spin, neglecting the contribution from bosonic fields to the spin tensor. Under this condition, we simply have $T^{\alpha}{}_{\beta\gamma}=K^{\alpha}{}_{\beta\gamma}$. We will consider this simplified regime later in some applications.

In general, from the contortion contributions to the Dirac Lagrangian (\ref{DiracLag}) only the (completely) antisymmetric part survives, giving:
\begin{equation}
\mathcal{L}_{\rm Dirac}=\tilde{\mathcal{L}}_{\rm Dirac}+3\breve{T}^{\lambda}\breve{s}^{D}_{\lambda} \,,
\end{equation}
where the axial vector part of torsion, in the full regime, reads
\begin{eqnarray}
\label{axialtorsion}
\breve{T}^{\lambda}&\equiv &\dfrac{1}{6}\epsilon^{\lambda\alpha\beta\gamma}T_{\alpha\beta\gamma}  \\
&=&\kappa^{2}\left[-\dfrac{\breve{s}^{\lambda}}{2}+\dfrac{\lambda}{6}\epsilon^{\mu\beta\gamma\lambda}\left(2\kappa^{2}s^{D}_{\rho[\mu\beta}A_{\gamma]}A^{\rho}+A_{[\mu}\tilde{F}_{\beta\gamma]}\right)\right], \nonumber
\end{eqnarray}
and is computed from Eq.(\ref{finalnewCartan}), and we have omitted the $D$ symbol in the Dirac axial spin vector to shorten notation. We therefore arrive at
\begin{eqnarray}
\label{newDiracLag}
\mathcal{L}_{\rm D}&=&\tilde{\mathcal{L}}_{\rm D}-\breve{s}_{\lambda}\breve{s}^{\lambda}\left(\dfrac{3\kappa^{2}}{2}+\lambda\kappa^{4}A^{2} \right)+\lambda\kappa^{4}(A\cdot \breve{s})^{2}
\nonumber \\
&&+\dfrac{\lambda\kappa^{2}}{2}\epsilon^{\mu\beta\gamma\lambda}\breve{s}_{\lambda}A_{[\mu}\tilde{F}_{\beta\gamma]} \ ,
\end{eqnarray}
where we have introduced the notation $\breve{s}^{2}\equiv \breve{s}^{\lambda}\breve{s}_{\lambda}$, $A^{2}\equiv A^{\lambda}A_{\lambda}$ and $\breve{s}\cdot A\equiv \breve{s}^{\lambda}A_{\lambda}$.
The first term of Eq.(\ref{newDiracLag}) is Dirac's Lagrangian on a (pseudo) Riemann spacetime, while the other terms come from the corrections of a RC geometry, including spin-spin self-interactions and non-minimal couplings with the bosonic fields.

As for the generalized Maxwell Lagrangian, in the regime of random fermionic spin distributions (zero average, macroscopic spin), where we retain only the terms quadratic with the Dirac spin quantities, we obtain
\begin{eqnarray}
\label{BosonLagcorr}
\mathcal{L}^{\rm M}_{\rm corr}&\approx&\lambda^{2}\kappa^{2}A^{[\mu}\tilde{F}^{\nu]\lambda}\tilde{F}_{\mu\nu}A_{\lambda}
	\nonumber \\
&&+\dfrac{2\lambda\kappa^{2}\tilde{F}_{\mu\nu}}{2+\lambda\kappa^{2}A^{2}}A^{[\mu}\tilde{s}^{\nu]}(1-\lambda \kappa^{2}A^{2})\nonumber \\
&&+\lambda^{3}\kappa^{4}A^{[\mu}\tilde{F}^{\nu]\lambda}A_{[\mu}\tilde{F}_{\nu]\gamma}A_{\lambda}A^{\gamma}\\
&&+\dfrac{4\lambda\kappa^{4}A^{[\mu}\tilde{s}^{\nu]}A_{[\mu}\tilde{s}_{\nu]}}{(2+\lambda\kappa^{2}A^{2})^{2}}\left[1-\lambda\kappa^{2}A^{2}(2-\lambda\kappa^{2}A^{2})\right]\nonumber \\
&&-\dfrac{\lambda\kappa^{4}}{2}\left[A^{2}\breve{s}^{2}-(A\cdot\breve{s})^{2}\right] \ . \nonumber
\end{eqnarray}
The first four terms correspond to self-interactions while the last one depends on the spinors via the Dirac axial vector $\breve{s}^{\lambda}$, and represents non-minimal boson-fermion interactions.

Let us note that in the simplified regime of fermionic torsion the Dirac Lagrangian boils down to
 \begin{equation}
\label{newDiracLagsimple}
\mathcal{L}_{\rm D}=\tilde{\mathcal{L}}_{\rm D}-\dfrac{3\kappa^{2}}{2}\breve{s}^{2} \ ,
\end{equation}
and the electromagnetic one to
\begin{equation}
\label{BosonLagsimple}
\mathcal{L}_{\rm M}=\tilde{\mathcal{L}}_{\rm M}-\lambda\left[ \dfrac{\kappa^{4}}{2}\left(\breve{s}^{2}A^{2}-(\breve{s}\cdot A)^2\right)-\dfrac{\kappa^2}{2}f^{\nu}\breve{s}_{\nu}\right] \ ,
\end{equation}
where we have introduced the (axial) vector
\begin{equation}
\label{smallf}
f^{\rho}\equiv \epsilon^{\lambda\mu\nu\rho}A_{\lambda}\tilde{F}_{\mu\nu} \ ,
\end{equation}
and the corresponding term can be neglected under the assumption of the random spin distribution.

From the matter Lagrangian in (\ref{MatterLag}), using (\ref{newDiracLag}) and(\ref{BosonLagcorr}), as well as (\ref{newDiracLagsimple}) and (\ref{BosonLagsimple}), the corresponding dynamical energy-momentum tensors which enter in the gravitational equations, can be obtained, as well as the matter fields equation of motion.  
\subsection{Gravitational field equations}

To cast the effective Einstein equations (\ref{eq:EOMeff}) under suitable form, we first note that the Ricci scalar of a RC spacetime geometry is related to the Riemann one in terms of the expression
\begin{equation}
R=\tilde{R}-2\tilde{\nabla}^{\lambda}K^{\alpha}{}_{\lambda\alpha} +g^{\beta\nu}\left(K^{\alpha}{}_{\lambda\alpha}K^{\lambda}{}_{\beta\nu}-K^{\alpha}{}_{\lambda\nu}K^{\lambda}{}_{\beta\alpha}\right) \ .
\nonumber
\end{equation}
One can then compute the torsion-induced piece of the effective energy-momentum tensor, $U_{\mu\nu}=-\frac{2}{\sqrt{-g}} \frac{\delta (\sqrt{-g}C)}{\delta g^{\mu\nu}}$, from the definition
\begin{equation} \label{eq:Cexp}
C\equiv-\dfrac{1}{2\kappa^2}
\Big(K^{\alpha\lambda}_{\;\;\;\;\alpha}K_{\;\;\lambda\beta}^{\beta}+K^{\alpha\lambda\beta}K_{\lambda\beta\alpha}\Big) \ .
\end{equation}
This yields quadratic corrections in torsion, $U\sim\kappa^{-2}T^{2}$, or in the spin variables, $U\sim \kappa^2 s^2$, via Cartan's equations. On the other hand, the dynamical energy-momentum tensor, $T_{\mu\nu}=-\frac{2}{\sqrt{-g}} \frac{\delta (\sqrt{-g}\mathcal{L}_m)}{\delta g^{\mu\nu}}$, is computed as usual from the matter Lagrangian, which yields the explicit result
\begin{equation} \label{eq:Tmunudym}
T_{\mu\nu}=\tilde{T}_{\mu\nu}-4j_{(\mu}A_{\nu)}+j^{\lambda}A_{\lambda}g_{\mu\nu}+\Pi^{M{\rm int}}_{\mu\nu}+\Xi^{D{\rm int}}_{\mu\nu}\,,
\end{equation}
 where
 $\tilde{T}_{\mu\nu}= \tilde{T}^{\rm{Dirac}}_{\mu\nu}+\tilde{T}^{\rm{Max}}_{\mu\nu}$, is the energy-momentum tensor for the matter fields in a Riemannian spacetime. In the above expression, the term
$\Pi^{M{\rm int}}_{\mu\nu}=-\dfrac{2}{\sqrt{-g}}\dfrac{\partial {\cal L}^{\rm M}_{\rm corr}}{\partial g^{\mu\nu}}$
arises from the (second term in the) bosonic Lagrangian (\ref{newMaxLagra}) and
includes non-minimal boson-fermion interactions (induced by torsion) and also bosonic self-interactions. Similarly, $\Xi^{D{\rm int}}_{\mu\nu}$ comes from the Dirac Lagrangian, i.e, $\Xi^{D{\rm int}}_{\mu\nu}=-\dfrac{2}{\sqrt{-g}}\dfrac{\partial {\cal L}_{\rm corr}^{\rm D}}{\partial g^{\mu\nu}}$, where $L_{\rm corr}^{\rm D}\equiv3\breve{T}^{\lambda}\breve{s}^{\rm D}_{\lambda}$, and it corresponds to non-minimal fermion-boson interactions (induced by torsion) and also spin-spin fermionic self-interactions.

To illustrate these expressions, let us consider the ansatz $\tilde{F}_{\mu\nu}=0$, corresponding to a spacetime with $A_{\mu}=(\phi(t),0,0,0)$). In this case we find $\tilde{s}^{\alpha\beta\gamma}=0$, and the last two terms in Eq.(\ref{eq:Tmunudym}) read
\begin{eqnarray}
\label{dynamicalenergymom}
\Pi^{M{\rm int}}_{\mu\nu}&+&\Xi^{D{\rm int}}_{\mu\nu}=
6\left( \kappa^{2}+\lambda\kappa^{4}A^{2} \right)\breve{s}_{\mu}\breve{s}_{\nu}+6\lambda\kappa^{4}\breve{s}^{2}A_{\mu}A_{\nu}
\nonumber \\
&-&16\lambda\kappa^{4}(A\cdot\breve{s})A_{(\mu}\breve{s}_{\nu)}
	\nonumber\\
&+&\dfrac{1}{2}\left[\lambda\kappa^{4}(A\cdot\breve{s})^{2}-\breve{s}^{2}\left(3\kappa^{2}+\lambda\kappa^{4}A^{2} \right)
\right]g_{\mu\nu},
\end{eqnarray}
while Eq.(\ref{eq:Cexp}) yields (by substituting the torsion components by spin quantities using Cartan's equations)
\begin{equation}
C=-\dfrac{\kappa^{2}}{2}\left[s^{\lambda}s_{\lambda}+ s^{\mu\nu\lambda}\left(s_{\nu\lambda\mu}+s_{\lambda\mu\nu}+s_{\mu\lambda\nu}\right)\right] \ .
\end{equation}
From this expression we can compute the torsion-induced contribution to the energy-momentum tensor (due to the Ricci scalar in a RC spacetime) as
\begin{eqnarray}
\label{Utensor}
U_{\mu\nu}&=&\kappa^{2}\Big\{A_{\mu}A_{\nu}2\lambda\kappa^{2}\Big[\breve{s}^{2}(2-\lambda\kappa^{2}A^{2})
	\nonumber \\
&&-\lambda\kappa^{2}\left[A^{2}\breve{s}^{2}-(A\cdot\breve{s})^{2}\right] \Big]
	 \\
&&+\breve{s}_{\mu}\breve{s}_{\nu}\left[2\lambda\kappa^{2}A^{2}(2-\lambda\kappa^{2}A^{2})-3\right]\Big\}
	\nonumber \\
	&&+4\lambda\kappa^{4}(2-\lambda\kappa^{2}A^{2})(A\cdot\breve{s})A_{(\mu}\breve{s}_{\nu)}+Cg_{\mu\nu}  \ , \nonumber
\end{eqnarray}
with
\begin{equation}
\label{Cterm}
C=-\dfrac{\kappa^{2}}{2}\left[\lambda\kappa^{2}(2-\lambda\kappa^{2}A^{2})(A^{2}\breve{s}^{2}-(A\cdot\breve{s})^{2})-\dfrac{3}{2}\breve{s}^{2}\right] \ .
\end{equation}
The effective energy-momentum tensor is therefore given by the expression
\begin{eqnarray}
\label{effectiveenergym}
T^{\rm eff}_{\mu\nu}&=&\tilde{T}_{\mu\nu}-4j_{(\mu}A_{\nu)}+j^{\lambda}A_{\lambda}g_{\mu\nu}\nonumber \\
&&-\breve{s}_{\mu}\breve{s}_{\nu}\kappa^{2}\left[ -3-\lambda\kappa^{2}A^{2}(1+2(\lambda\kappa^{2}A^{2}))\right]  \nonumber \\
&&+ A_{\mu}A_{\nu}\lambda\kappa^{4}\Big[\breve{s}^{2}[6+2(2-\lambda\kappa^{2}A^{2})]
	\nonumber \\
&&-2\lambda\kappa^{2}[A^{2}\breve{s}^{2}-(A\cdot\breve{s})^{2}]\Big]
\nonumber \\
&&+ A_{(\mu}\breve{s}_{\nu)}(A\cdot \breve{s})\lambda\kappa^{4}\left(4(2-\lambda\kappa^{2}A^{2})-16\right) \nonumber \\
 &&-\Big[\kappa^{2}\breve{s}^{2}\Big(\dfrac{3}{4}+\dfrac{\lambda\kappa^{2}A^{2}}{2}\Big)
	\nonumber \\
&&+\dfrac{\lambda\kappa^{4}}{2}\big[(2-\lambda\kappa^{2}A^{2})(A^{2}\breve{s}^{2}-(A\cdot\breve{s})^{2}) \nonumber \\
&&-(A\cdot\breve{s})^{2}\big] \Big]g_{\mu\nu} \,.
\end{eqnarray}

Another simplifying scenario is when torsion exclusively results from the spin tensor of fermions, neglecting the bosonic contribution to the Cartan equations. Furthermore, let us keep only terms quadratic in the Dirac spin variables, with the linear ones vanishing upon averaging for random distributions of spin. Under these assumptions, using Cartan's equations (\ref{fermionictorsion}) we get
\begin{eqnarray}
\label{Utensor2}
U_{\mu\nu}&=&\kappa^{2}\left(\dfrac{3}{4}\breve{s}^{2}g_{\mu\nu}-3\breve{s}_{\mu}\breve{s}_{\nu}\right) \ ,
\end{eqnarray}
and
\begin{eqnarray}
\label{dynamicalenergymomDirac}
\Xi^{D}_{\mu\nu}&=&
\kappa^{2}\left(6\breve{s}_{\mu}\breve{s}_{\nu}-\dfrac{3}{2}\breve{s}^{2}g_{\mu\nu}\right) \ ,
\end{eqnarray}
(derived from Eq.(\ref{newDiracLagsimple}))
and
\begin{eqnarray}
\label{dynamicalenergymomMax}
\Pi^{M}_{\mu\nu}&=&\lambda\kappa^{4}\Big[\breve{s}^{2}A_{\mu}A_{\nu}+A^{2}\breve{s}_{\mu}\breve{s}_{\nu}-4(\breve{s}\cdot A)\breve{s}_{(\mu}A_{\nu)}
\nonumber \\
&&-\dfrac{g_{\mu\nu}}{2}\Big(A^{2}\breve{s}^{2}-(A\cdot \breve{s})^{2}\Big)\Big] \ ,
\end{eqnarray}
(derived from (\ref{BosonLagsimple})). Thus, the final result reads
\begin{eqnarray}
\label{effectiveenergym2}
T^{\rm eff}_{\mu\nu}&=&\tilde{T}_{\mu\nu}-4j_{(\mu}A_{\nu)}+j^{\lambda}A_{\lambda}g_{\mu\nu}
	-\kappa^{2}\breve{s}_{\mu}\breve{s}_{\nu}\left(-3+\lambda\kappa^{2}A^{2}\right)
	\nonumber \\
&&+\lambda\kappa^{4}\left[A_{\mu}A_{\nu}\breve{s}^{2}- 4A_{(\mu}\breve{s}_{\nu)}(A\cdot \breve{s})\right]
	\\
&&-\left[\dfrac{3\kappa^{2}}{4}\breve{s}^{2}+\dfrac{\lambda\kappa^{4}}{2}\left(A^{2}\breve{s}^{2}-(A\cdot\breve{s})^{2}\right)\right]g_{\mu\nu} \ .  \nonumber
\end{eqnarray}

\section{Cosmological dynamics} \label{secIV}

The full set of gravitational, electromagnetic and fermionic equations of the ECDM model can be derived from the action (\ref{eq:actionEC}) with the matter Lagrangian (\ref{MatterLag}) which implement the minimal coupling between torsion and matter fields. This results in torsion-induced non-minimal couplings between fermions and bosons and also in self-interactions \cite{Cabral:2019gzh}. In this section we shall study the cosmological dynamics associated to this framework, under the assumption of randomly oriented spin densities (unpolarized matter).

\subsection{Fluid description and Friedman equations}\label{secIVA}

Let us assume a homogeneous and isotropic Universe, which is described by the Friedman-Lema\^{i}tre-Robertson-Walker (FLRW) metric, given by the line element
\begin{equation} \label{eq:FLRW}
ds^2=dt^2-a^2(t)\left(\frac{dr^2}{1-kr^2}+r^2 d\Omega^2\right)\,,
\end{equation}
where $a(t)$ is the scale factor and $k$ denotes the curvature of space. As usual, matter is described by a perfect fluid with an energy-momentum tensor ${T^\mu}_{\nu}=\text{diag}(\rho,-p,-p,-p)$, where $\rho$ and $p$ are the energy density and pressure, respectively. For the sake of this section we shall consider both (relativistic) fermionic matter and radiation coupled to spacetime torsion.

One of the most common approaches to cosmology with spin is to consider the Weyssenhof spin fluid (see Ref. \cite{Obukhov} for details), which can be seen as the classical approximation of a fluid of fermionic matter with macroscopic spin effects. In this paper, however, we shall take instead the approach from fundamental Dirac spinors. To this end, it is usual to consider that for comoving observers the spin (axial) vector is spatial, i.e., $\breve{s}^{\lambda}u_{\lambda}=0$, where $u^{\alpha}=(1,0,0,0)$ is the fluid's unit four-velocity field.
Nevertheless, since fermionic fields are appropriately represented by four-spinors, which can be regarded as fundamental quantum fields, in order to establish a (macroscopic) fluid description we will adopt the correspondence principle approach, through the definitions\footnote{One can see that
\begin{equation}
\breve{s}^{2}\equiv\dfrac{\hbar^{2}}{4}\left<\bar{\psi}\gamma^{\nu}\gamma^{5}\psi(\bar{\psi}\gamma_{\nu}\gamma^{5}\psi)\right>=\dfrac{\hbar^{2}}{4}\left<\bar{\psi}\gamma^{a}\gamma^{5}\psi(\bar{\psi}\gamma_{a}\gamma^{5}\psi)\right> \ , \nonumber
\end{equation}
should scale as $\breve{s}^{2}\sim \left<(\bar{\psi}\psi)^{2}\right>\sim n^{2}(t)$, where $n$ is the number density of fermions.
Here, $a=0,1,2,3$ and the usual constant Pauli-Dirac matrices $\gamma^{c}$, which obey $\left\lbrace \gamma^{a},\gamma^{b}\right\rbrace =2\eta^{ab}I$, are related to the $\gamma^{\mu}$ matrices via $\gamma^{\mu}e^{a}_{\;\;\mu}=\gamma^{a}$, where $e^{a}_{\;\;\mu}$ are the tetrads satisfying $g_{\mu\nu}=\eta_{ab}e^{a}_{\;\;\mu}e^{b}_{\;\;\nu}$ and $e^{a}_{\;\;\mu}e^{\;\;\mu}_{b}=\delta^{a}_{b}$, $e^{c}_{\;\;\nu}e^{\;\;\mu}_{c}=\delta^{\mu}_{\nu}$ and $\eta_{ab}$ is the Minkowski metric.}
\begin{equation}
\breve{s}^{2}\equiv\dfrac{\hbar^{2}}{4}\left<\bar{\psi}\gamma^{\nu}\gamma^{5}\psi(\bar{\psi}\gamma_{\nu}\gamma^{5}\psi)\right> \ .
\end{equation}
Accordingly, in the expressions for the effective energy densities and pressures all (fermionic) spin quantities should be regarded as expectation values.
Moreover, throughout the rest of this paper we shall assume that the cosmological fluid has vanishing macroscopic intrinsic spin on average ($\bar{\breve{s}}\simeq 0$), under the unpolarized matter assumption. However, quantities quadratic in spin do not average to zero. Thus, by taking $\overline{\breve{s}^{2}}=g^{kk}\overline{\breve{s}_{k}\breve{s}_{k}}$ and $\overline{\breve{s}^{i}\breve{s}_{j}}\approx {\rm diag}(\overline{\breve{s}_{1}\breve{s}_{1}}g^{11}\;\overline{\breve{s}_{2}\breve{s}_{2}}g^{22}\;\overline{\breve{s}_{3}\breve{s}_{3}}g^{33})\approx \delta^{i}_{j}\overline{\breve{s}^{2}}/3 \ , $ invoking isotropy, we assume that for fermions we have (on average)
\begin{equation}
\overline{\breve{s}^{2}}=\beta_{s} n^{2}(t),\qquad \overline{\breve{s}^{i}\breve{s}_{j}}=\dfrac{\overline{\breve{s}^{2}}}{3}\delta^{i}_{j}\sim \dfrac{n^{2}(t)}{3}\delta^{i}_{j} \ ,
\end{equation}
where $n(t)\sim a^{-3}$ and $\vert\beta_{s}\vert\sim\hbar^{2}$. We therefore neglect possible anisotropic pressure contributions from the $\breve{s}^{i}\breve{s}_{j}$ terms.

As for the bosonic vector potential, we use two different ansatze: i) $A_{\mu}=(\phi(t),0,0,0)$, therefore $\tilde{F}_{\mu\nu}=0$ and $\tilde{s}^{\alpha\beta\gamma}=0$, and ii) $A=(0,\vec{A}(t))$ (with its orientation randomly distributed to respect isotropy), therefore $\overline{\vec{A}}\simeq 0$ and we have $\overline{A^{k}A_{k}}=\overline{{\vec{A}}^{2}}\neq 0$ and take $\overline{A^{i}A_{j}}\approx\delta^{i}_{j}\overline{{\vec{A}}^{2}}/3$, again invoking isotropy.

\subsubsection{Friedman equations}

The generalized Einstein equations (\ref{eq:EOMeff}) can be written, in the FLRW background (\ref{eq:FLRW}), for isotropic pressure, as
\begin{eqnarray}
\left(\dfrac{\dot{a}}{a}\right)^2&=&\frac{\kappa^2}{3}\left(\rho+\rho^{\rm corr}\right)-\frac{k}{a^{2}}, \label{eq:F1} \\
\dfrac{\ddot{a}}{a}&=&-\frac{\kappa^2}{6}\left[3(p+p^{\rm corr})+(\rho+\rho^{\rm corr})\right],
\label{eq:F2}
\end{eqnarray}
where as usual dots over functions mean time derivatives. Here $\rho^{\rm eff}=\rho+\rho^{\rm corr}$ and $\rho^{\rm corr}=\rho^{s}+\rho^{s-A}+\rho^{A-A}$ with the corrections to GR corresponding to the spin-spin interaction energy, the non-minimal interactions between fermionic spin and the bosonic four-potential, as well as bosonic self-interactions, and the same apply to the pressure contributions. We shall split now our analysis in four different cases.\\

$\bullet$ \underline{CASE I}: Under the ansatz of random fermionic spin
and $A_{\mu}=(\phi(t),0,0,0)$, using Eq.(\ref{effectiveenergym}) we get for the correction densities and pressures:
\begin{eqnarray}
\rho^{\rm corr}&\simeq&-\kappa^{2}\breve{s}^{2}\left[\dfrac{3}{4} +\lambda\kappa^{2}\phi^{2}\left(\dfrac{7}{2}\lambda\kappa^{2}\phi^{2}-\dfrac{17}{2} \right) \right] ,  \\
p^{\rm corr}\delta^{i}_{j}&\simeq&-\kappa^{2}\breve{s}^{2}\left[\dfrac{1}{4} +\lambda\kappa^{2}\phi^{2}\left(\dfrac{1}{6}-\dfrac{1}{6}\lambda\kappa^{2}\phi^{2}\right) \right]\delta^{i}_{j} \ ,
\end{eqnarray}
respectively.\\

$\bullet$ \underline{CASE II}: Under the ansatz of $A_{\mu}$ to be spatial and randomly oriented, $A=(0,\vec{A}(t))$, the Maxwell Lagrangian in (\ref{BosonLagcorr}) can be written as
\begin{equation}
\mathcal{L}^{\rm M}_{\rm corr}=f(A)\tilde{F}^{0}_{k}\tilde{F}_{0j}A^{k}A^{j}-\dfrac{\lambda\kappa^{4}}{2}\left[A^{2}\breve{s}^{2}-(A\cdot\breve{s})^{2}\right] \ .
\end{equation}
Using now Eq.(\ref{newDiracLag}), we obtain
\begin{equation}
\rho^{\rm corr}=-\kappa^{2}\breve{s}^{2}\Big(\dfrac{3}{2}+\lambda\kappa^{2}A^{2}\Big) -f(A)\tilde{F}_{0k}\tilde{F}_{0j}A^{k}A^{j}+\rho^{\rm corr}_{U}   ,
\end{equation}
with $f(A)$ given by
\begin{eqnarray}
f(A)&=&\dfrac{\lambda^{2}\kappa^{2}}{2}\Big[\dfrac{\lambda\kappa^{2}A^{2}(4+\lambda\kappa^{2}A^{2}(\lambda\kappa^{2}A^{2}-1))+2}{(2+\lambda\kappa^{2}A^{2})^{2}} 
\nonumber \\
&&+\lambda\kappa^{2}A^{2}-2\Big] \ .
\end{eqnarray}
Here $\rho^{\rm corr}_{U}$ is the contribution coming from the $U_{\mu\nu}$ tensor, which gives a quite cumbersome and far from illuminating expression, and the other terms come from the corrections to the $T_{\mu\nu}$ tensor. Since the relevant torsion-induced corrections coming from the tensor $U_{\mu\nu}$ are quadratic in the contortion, by taking into account Eq.(\ref{finalnewCartan}) and neglecting terms  that scale linearly with $\breve{s}$, we obtain (approximately) a similar expression
\begin{equation}
\rho^{\rm corr}\approx\breve{s}^{2}\left(C+(h+bA^{2})A^{2}\right)+ h(A)\dot{A}_{j}\dot{A}_{k}A^{j}A^{k} \ ,
\end{equation}
with $h(A)$ some expression of $A$ with dimensions of $\lambda^{2}\kappa^{2}$. The first term includes spin-spin fermion self-interactions and fermion-boson non-minimal couplings, while the last term represents the energy density from bosonic self-interactions, although other self-interactions of the form $\sim\lambda^{2}\kappa^{4}x(A)\tilde{F}^{2}A^{6}$ can also be present.

As for the pressure corrections, $p^{\rm corr}$, neglecting anisotropic stresses we arrive at a similar (approximate) expression
\begin{equation}
p^{\rm corr}\delta^{k}_{m}\approx\left[\breve{s}^{2}\left(D+(q+cA^{2})A^{2}\right)+t(A)\dot{A}_{j}\dot{A}_{k}A^{j}A^{k}\right]\delta^{k}_{m}  \ .
\end{equation}
The anisotropic stresses are present, in general, coming for instance from a term of the form $\tilde{F}_{0k}A^{k}\tilde{F}_{0(i}A_{j)}$ in the effective energy momentum tensor, which can be written as $\sim\tilde{F}_{0i}\tilde{F}_{0j}A^{2}$ using $\overline{A^{i}A_{j}}\approx\overline{{\vec{A}}^{2}}\delta^{i}_{j}/3$. The corresponding stresses $T^{k}_{j}$ can be recast into the (averaged) isotropic form $\sim\dot{A}^{m}\dot{A}_{m}A^{2}\delta^{k}_{j}$, by making the approximation $\overline{\dot{A}^{k}\dot{A}_{j}}\approx\overline{\dot{A}^{m}\dot{A}_{m}}\delta^{k}_{j}/3$. In this case, the final expression would be approximately isotropic, having exactly the same functional form as in the equation above. The second term can be simplified, using again $\overline{A^{i}A_{j}}\approx\overline{{\vec{A}}^{2}}\delta^{i}_{j}/3$ and $\overline{\dot{A}^{k}\dot{A}_{j}}\approx\overline{\dot{A}^{m}\dot{A}_{m}}\delta^{k}_{j}/3$, which yields
\begin{equation}
\rho^{\rm corr}\approx\breve{s}^{2}\left(C+(h+bA^{2})A^{2}\right)+\frac{1}{9}h(A)\dot{A}^{2}A^{2} \ ,
\end{equation}
and
\begin{equation}
p^{\rm corr}\delta^{k}_{m}\approx\left[\breve{s}^{2}\left(D+(q+cA^{2})A^{2}\right)+\frac{1}{9}t(A)\dot{A}^{2}A^{2}\right]\delta^{k}_{m}  \ .
\end{equation}

$\bullet$ \underline{CASE III} (fermionic torsion):
Let us now consider the regime in which the bosonic spin tensor does not contribute to torsion, i.e., bosonic fields are influenced by spacetime torsion and affect the cosmological dynamics but do not back-react on torsion. In this case, using the ansatz  $A_{\mu}=(\phi(t),0,0,0)$, from Eq.(\ref{effectiveenergym2}) we find
\begin{eqnarray}
\rho^{\rm corr}&\simeq&-\kappa^{2}\breve{s}^{2}\left(\dfrac{3}{4} +\dfrac{\lambda\kappa^{2}}{2}\phi^{2}\right)  \ ,  \\
p^{\rm corr}\delta^{i}_{j}&\simeq&-\kappa^{2}\breve{s}^{2}\left(\dfrac{1}{4}-\dfrac{5\lambda\kappa^{2}}{6}\phi^{2}\right) \delta^{i}_{j}  \ .
\end{eqnarray}

$\bullet$ \underline{CASE IV} (fermionic torsion): Under the ansatz $A_{\mu}=(0,\vec{A}(t))$ we get
\begin{eqnarray}
\rho^{\rm corr}&\simeq &-\kappa^{2}\breve{s}^{2}\left(\dfrac{3}{4} +\dfrac{\lambda\kappa^{2}}{3}\vec{A}^{2}\right)  \ , \\
p^{\rm corr}\delta^{i}_{j}&\simeq &-\kappa^{2}\breve{s}^{2}\left(\dfrac{1}{4}-\dfrac{2\lambda\kappa^{2}}{3}\vec{A}^{2}\right)\delta^{i}_{j}  \ .
\end{eqnarray}
A slight modification of this case occurs when, instead of the approximations $\overline{A^{k}A_{j}}\approx\overline{A^{2}}\delta^{k}_{j}/3$, and $\overline{\breve{s}^{k}\breve{s}_{j}}\approx\overline{\breve{s}^{2}}\delta^{k}_{j}/3$, we consider $\overline{A^{k}A_{j}}\approx\overline{A^{2}}\delta^{k}_{j}$, and $\overline{\breve{s}^{k}\breve{s}_{j}}\approx\overline{\breve{s}^{2}}\delta^{k}_{j}$. This way we arrive at the following expressions:
\begin{eqnarray}
\rho^{\rm corr} &\simeq &-\kappa^{2}\breve{s}^{2}\left(\dfrac{3}{4}-\lambda\kappa^{2}\vec{A}^{2}\right)  \ , \\
p^{\rm corr}\delta^{i}_{j}&\simeq&-\kappa^{2}\breve{s}^{2}\left(\dfrac{9}{4}+\lambda\kappa^{2}\vec{A}^{2}\right)\delta^{i}_{j} \ .
\end{eqnarray}

In all these cases we used $\breve{s}^{2}=\beta_{s} n^{2}(t)=\alpha_{s} a^{-6}$, which means that in the very early universe the spin-spin effects start to strongly dominate over the usual energy density and pressure of the relativistic fluid. The $\breve{s}^{2}\sim a^{-6}$ behaviour is usually considered in cosmological applications of ECSK theory for fluids with spin. It follows directly from a conserved fluid component corresponding to the spin-spin interaction, with an effective stiff-like equation of state, $w^{s}=p^{s}/\rho^{s}=1$. It is also a natural result from the theory of fermionic Dirac spinors. In ECSK theory it is the negative value of $\rho_{s}$ that acts as a repulsive effect. In the present ECDM model the other contributions ($\rho_{s-A}$) may affect the early Universe dynamics by reinforcing or counter-acting this repulsive phenomena, depending on the sign and strength of these extra terms.

In order to explore the solutions of the dynamics in this torsion era we need to evaluate the time dependence of the bosonic four-potential, or equivalently its behaviour with the cosmological scale factor.
Besides the Friedman equations we have at our disposal also the effective energy-momentum conservation equation, $\tilde{\nabla}_{\mu}T^{\mu\nu}_{\rm eff}=0$, the generalized electromagnetic equations and the corresponding effective charge conservation. Beyond the fluid approach, one needs to consider the dynamics of fundamental fermionic degrees of freedom, that is, the Dirac equation in the FLRW cosmological framework.

\subsubsection{Effective conservation equation}

Let us thus consider the generalized energy-momentum conservation:
\begin{equation}
\dot{\rho}_{\rm eff}+3H(\rho_{\rm eff}+p_{\rm eff})=0  \ ,
\end{equation}
with $\rho_{\rm eff}=\rho+\rho_{\rm corr}$ and $p_{\rm eff}=p+p_{\rm corr}$. For simplicity, we shall consider the different contributions to the effective energy density as different fluid components which are independently conserved. These components correspond to the usual relativistic fluid (``radiation'') term, the torsion-induced spin-spin interaction, an additional term representing the non-minimal interaction between fermionic spin and the bosonic potential, as well as bosonic self-interactions (both also induced by the spin-torsion Cartan relation), i.e.,
\begin{equation}
\rho_{\rm eff}=\rho+\rho^{s-s}+\rho^{s-A}+\rho^{A-A} \ ,
\end{equation}
and analogously for the pressures. From now on we will focus our attention in cases I, III and IV, neglecting in this way the bosonic self interactions $\rho^{A-A}$. Therefore, independent conservation implies
\begin{equation}
\dot{\rho}_{s-A}+3H(\rho_{s-A}+p_{s-A})=0 \ .
\end{equation}
This can be solved in order to provide the $A_{\mu}(t)$ dependence or, alternatively, to get the dependence with the scale factor $A(a)$ as
\begin{equation}
\dfrac{d\rho_{s-A}}{da}+\dfrac{3}{a}(w_{s-A}+1)\rho_{s-A}=0  \ ,
\end{equation}
which yields the solution $\rho_{s-A}\sim a^{-3(w_{s-A}+1)}$ for constant $w_{s-A}$.

\subsubsection{Torsion due to fermionic spin, neglecting the contribution from the bosonic spin tensor}

As a specific example let us consider the Cases III and IV above. We have $w^{s-A}=-5/3$ and $w^{s-A}=-2$, respectively and, therefore, $\rho^{s-A}\sim a^{2}$ and $\rho^{s-A}\sim a^{3}$, respectively, which in turn implies that $\phi\sim a^{4}$ and $A_{j}A^{j}=(A_{j})^{2}g^{jj}\sim a^{9}$, respectively. In the last case, since $g^{jj}\sim a^{-2}$ we get $A_{j}\sim a^{11/2}$.  More rigorously, for $\rho_{s-A}=C\breve{s}^{2}\phi^{2}$, (as in Case III) with $C$ a constant and $p_{s-A}/\rho_{s-A}=w_{s-A}$ also constant, we obtain
\begin{equation}
\breve{s}^{2}\dfrac{d\phi^{2}}{da}+\dfrac{3}{a}\left(w_{s-A}+1+\dfrac{a}{3\breve{s}^{2}}\dfrac{d\breve{s}^{2}}{da}\right)\breve{s}^{2}\phi^{2}=0  \ ,
\end{equation}
which yields the solution
\begin{equation}
\phi(a)\sim a^{-3(w_{s-A}-1)/2}  \ ,
\end{equation}
This is compatible with the previous conclusion that for  $w_{s-A}=-5/3$ we get $\phi\sim a^{4}$.
Analogously, for $\rho_{s-A}=C\breve{s}^{2}\vec{A}^{2}$ (as in Case IV) with $p_{s-A}/\rho_{s-A}=w_{s-A}$ constant, we obtain
\begin{equation}
\vec{A}^{2}\sim a^{-3(w_{s-A}-1)}  \ ,
\end{equation}
and therefore
\begin{equation}
A_{j}^{2}\sim a^{-3(w_{s-A}+1-2)+2}  \ ,
\end{equation}
which for $w_{s-A}=-2$, provides $A_{j}\sim a^{11/2}$.

Let us summarize the main conclusions so far. Under the simplifying assumption that the energy contributions from the masses of relativistic fermions and bosons, from the spin-spin interaction and from the fermion-boson non-minimal interactions are separately conserved, with no energy exchanges between them, the terms representing the non-minimal interactions scale with $\rho^{s-A}\sim-\lambda\kappa^{4}\hbar^{2}a^{2}$ or $\rho^{s-A}\sim\lambda\kappa^{4}\hbar^{2}a^{3}$ depending on the ansatz for the bosonic four-potential. In the alternative derivation of Case IV we get instead $\rho^{s-A}\sim-\lambda\kappa^{4}\hbar^{2}a^{0}$ ($w^{s-A}=-1$). This means that at least when torsion is exclusively due to fermionic spin, the non-minimal couplings induced by the $U(1)$ symmetry breaking should not introduce major deviations from the usual ECSK theory in the torsion era of the early Universe. This follows from the $\rho^{s}\sim a^{-6}$ behaviour that dominates the early-Universe dynamics. However interesting late-time effects can occur, as we shall see.

\subsubsection{Torsion due to the spin tensor of fermions and bosons}

In this scenario, for Case I we have
\begin{eqnarray}
\rho^{s-A}&=&C\breve{s}^{2}\phi^{2}(h+b\phi^{2})   \ , \\
p^{s-A}&=&C\breve{s}^{2}\phi^{2}(d+c\phi^{2}) \ .
\end{eqnarray}
Assuming that $w_{s-A}(a)=p^{s-A}/\rho^{s-A}\simeq {\rm constant}$ we get
\begin{equation}
\rho^{s-A}\sim a^{-3(w_{s-A}+1)}  \ .
\end{equation}
Moreover, in this case we can take the approximation $w_{s-A}(a)\simeq c/b= -1/24$, that gets progressively more accurate for larger values of $\phi$, and we have
\begin{equation}
\rho^{s-A}\sim O(a^{-2,88})  \ ,
\end{equation}
again not competing with the $a^{-6}$ behaviour of the spin-spin energy density. The evolution for $\phi(a)$ can be then inferred from
\begin{equation}
\phi^{2}(h+b\phi^{2}) \sim O\left( a^{-3(w_{s-A}-1)}\right)  \ ,
\end{equation}
which implies that $\phi\sim O(a^{0,78})$ or, alternatively, from the conservation equation, leading to
\begin{equation}
 \dfrac{d\phi^{2}}{da}+\dfrac{3}{a}\left(w_{s-A}+1+\dfrac{a}{3\breve{s}^{2}}\dfrac{d\breve{s}^{2}}{da}\right)\phi^{2}+\dfrac{b\phi^{2}}{(h+b\phi^{2})}\dfrac{d\phi^{2}}{da}=0 ,
\end{equation}
which yields the solution
\begin{equation}
a(\phi)\sim \exp\left[{\dfrac{1}{3\bar{w}\phi}+\sqrt{\frac{b}{h}}\tan^{-1}\left(\sqrt{\frac{b}{h}}\,\phi\right)}\right]  \ ,
\end{equation}
with $\bar{w}\equiv w_{s-A}-1$.

More rigorously, if we do not assume $w_{s-A}(a)=p^{s-A}/\rho^{s-A}$ to be constant then we get
\begin{eqnarray}
\dfrac{d\phi^{2}}{da} \left[1+\dfrac{b\phi^{2}}{(h+b\phi^{2})}\right]+\dfrac{3}{a}\left(\dfrac{d+c\phi^{2}}{h+b\phi^{2}}\right)\phi^{2}
	\nonumber \\
=
-\dfrac{3}{a}\left(1+\dfrac{a}{3\breve{s}^{2}}\dfrac{d\breve{s}^{2}}{da}\right)\phi^{2}  \ ,
\end{eqnarray}
which yields the following solution
\begin{eqnarray}
&&a(\phi)\sim  \exp \Bigg\{-\dfrac{h}{3(h-d)\phi}
	\nonumber \\
&&+\dfrac{[b(h-2d)+ch]\tan^{-1}[(\sqrt{b-c}\phi)/\sqrt{h-d}]}{3\sqrt{b-c}(h-d)^{3/2}} \Bigg\} \ .
\end{eqnarray}
For the values of $h,b,d,c$ given in the expression of Case I, we obtain a specific bi-parametric family of curves (depending on the parameter $\lambda$ and an integration constant), which show $\phi$ increasing with increasing scale factor in the domains where the function is invertible. We obtain a similar solution for Case II, with $\rho^{s-A}\approx C\breve{s}^{2}A^{2}(h+bA^{2})$ if we neglect the $\sim(\dot{A}_{j}A^{j})^{2}$ term by replacing $\phi\rightarrow A$ and $h,b$ by the corresponding coefficients.

As a final comment, let us mention that, as usual, the cosmological solutions for the evolution of the scale factor can be derived from the expression ($da/d\eta=a^{2}H$)
\begin{equation}
\int\dfrac{da}{a^{2}\left(\kappa^{2}\rho^{\rm eff}(a)/3-k/a^{2}\right)^{1/2}}=\int d\eta + C,
\end{equation}
where $\eta$ is the usual conformal time, $dt=ad\eta$.

\subsection{Bouncing Cosmology }
\subsubsection{Non-singular solutions}

In principle, the minimum of the scale factor, which is present in the ECSK theory, should change in the ECDM model presented here. The Friedman equations can be combined as
\begin{equation} \label{eq:F3}
H^{2}(a)=\frac{\kappa^{2}}{3}\left[\alpha_{\rm rad}a^{-4}-\kappa^{2}\alpha_{s}a^{-6}+\rho^{s-A}(a)\right]-ka^{-2}  \ ,
\end{equation}
with $\rho^{s}(a)=-\kappa^{2}\alpha_{s}a^{-6}$, $\alpha_{s}>0$ and $\rho^{s-A}(a)=\lambda\rho^{s}f(A(a))$. By simplicity let us take the choice $k=0$, and by looking for the zeroes of $H^{2}(a)=0$ we get the equation
\begin{equation}
a^{2}-\dfrac{\kappa^{2}\alpha_{s}}{\alpha_{\rm rad}}+\dfrac{a^{6}\rho^{\rm s-A}(a)}{\alpha_{\rm rad}}=0  \ .
\end{equation}
In the standard ECSK theory ($\rho^{\rm s-A}$ switched off) we obtain the value of the scale factor at the bounce:
\begin{equation}
a_{\rm b}=\sqrt{\dfrac{\kappa^{2}\alpha_{s}}{\alpha_{\rm rad}}}  \ .
\end{equation}
For the ECDM model considered in this work, the exact value for  the scale factor at the bounce will depend on the parameters $\alpha_{r},\alpha_{s}$ as well as on the parameter $\lambda$ and on the value of $\rho^{\rm s-A}$ at some reference time.   We can take the general case with $\rho^{\rm s-A}=\alpha_{s-A}a^{b}$ and for the cases we have seen above (for instance $b=-2.88$, $b=2$ and $b=0$), the corresponding expressions for the scale factor at the bounce can be obtained.

To this end, let us consider first the Friedman equation without the (dust) matter term, which can be written as
\begin{equation}
H^{2}(x)=H^{2}_{0}\left(\Omega^{rad}_{0}x^{-4}+\Omega^{s}_{0}x^{-6}+\Omega^{s-A}_{0}x^{b}+\Omega^{k}_{0}x^{-2}\right)  \ ,
\end{equation}
with $x\equiv a/a_{0}$ and the parameters
\begin{eqnarray}
\alpha^{s}&=&-3\kappa^{-4}H^{2}_{0}\Omega^{s}_{0}a_{0}^{6}, \hspace{0.5cm}
\alpha^{rad}=3\kappa^{-2}H^{2}_{0}\Omega^{rad}_{0}a_{0}^{4}, \nonumber   \\
\alpha^{s-A}&=&3\kappa^{-2}H^{2}_{0}\Omega^{s-A}_{0}a_{0}^{-b}, \hspace{0.3cm} \Omega^{s-A}_{0}=\lambda f(A)\Omega^{s}_{0}(\frac{a}{a_{0}})^{-6-b} \ , \nonumber
\end{eqnarray}
and  $\vert\Omega^{s}_{0}\vert\sim\kappa^{2}\hbar^{2}\Omega^{mat}_{0}n_{0}$, with $n_{0}\sim(n_{0}/n^{\gamma CMB}_{0})n^{\gamma CMB}_{0}$ being the present fermion density number as a function of the ratio of fermions to CMB photons. In the expression $\Omega^{s-A}_{0}=\lambda f(A)\Omega^{s}_{0}(\frac{a}{a_{0}})^{-6-b}$ one can see that $f(A)\sim a^{6+b}$, which is compatible with $\rho^{s-A}(a)=\lambda\rho^{s}f(A)=\alpha^{s-A}a^{b}$. If we include now the matter term, for different values of $b$ (positive or negative) one gets  a bounce in the early universe just like in the usual ECSK cosmology, where the scale factor and the energy densities remain finite. To illustrate this idea, in the case $\rho^{\rm s-A}(a)\sim a^{-4}$  ($b=-4$) with $k=0$ we get
 \begin{equation}
a_{\rm b}=\sqrt{\dfrac{\kappa^{2}\alpha_{s}}{\alpha_{\rm rad}+\vert\alpha_{s-A}\vert}}  \ .
\end{equation}
For the specific case of spherical spatial hypersurfaces of constant cosmic time, $k=1$, we have the following two solutions
\begin{eqnarray}
a_{\rm b}&=&\Bigg[\dfrac{\kappa^{2}}{6}(\alpha_{\rm rad}+\vert\alpha_{s-A}\vert)
	\nonumber \\
&& \mp\dfrac{1}{6}\sqrt{-12\kappa^{2}\alpha_{s}+\kappa^{4}(-\vert\alpha_{s-A}\vert-\alpha_{\rm rad})^{2}}\Bigg]^{1/2}  \ .
\end{eqnarray}
Finally, for hyperbolic spatial hypersurfaces of constant cosmic time, $k=-1$, we arrive at the following two solutions
\begin{eqnarray}
&& a_{\rm b} = \Bigg[\dfrac{-\kappa^{2}}{6}(\vert\alpha_{s-A}\vert+\alpha_{\rm rad})
\nonumber \\
 && \qquad \mp  \frac{1}{6} \sqrt{12\kappa^{2}\alpha_{s}+\kappa^{4}(\alpha_{\rm rad}+\vert\alpha_{s-A}\vert)^{2}}  \Bigg]^{1/2}  \ .
\end{eqnarray}
These expressions can be compared with the corresponding solutions for the ECSK model: for $k=1$ we have
\begin{eqnarray}
a_{\rm b}= \sqrt{\dfrac{\kappa^{2}\alpha_{\rm rad}\mp\sqrt{\kappa^{4}\alpha_{\rm rad}^{2}-12\kappa^{2}\alpha_{s}}}{6}} \ ,
\end{eqnarray}
and for $k=-1$:
\begin{eqnarray}
a_{\rm b}= \sqrt{\dfrac{-\kappa^{2}\alpha_{\rm rad}\mp\sqrt{\kappa^{4}\alpha_{\rm rad}^{2}+12\kappa^{2}\alpha_{s}}}{6}} \ .
\end{eqnarray}

For ECDM theory with $b=-2$ we get similar expressions, for flat geometries, $k=0$:
\begin{equation}
a_{\rm b}=\sqrt{\dfrac{\alpha_{\rm rad}\pm\sqrt{\alpha_{\rm rad}^{2}-4\kappa^{2}\alpha_{s}\vert\alpha_{s-A}\vert}}{2\vert\alpha_{s-A}\vert}} \ , \nonumber
\end{equation}
two solutions for spherical geometries, $k=1$:
\begin{equation}
a_{\rm b}=\sqrt{\dfrac{\kappa^{2}\alpha_{\rm rad}\pm\sqrt{\kappa^{4}\alpha_{\rm rad}^{2}-12\kappa^{4}\alpha_{s}-4\kappa^{6}\alpha_{s}\vert\alpha_{s-A}\vert}}{6+2\kappa^{2}\vert\alpha_{s-A}\vert}} \ ,
\end{equation}
and also two solutions for the hyperbolic geometries, $k=-1$:
\begin{equation}
a_{\rm b}= \sqrt{\dfrac{\kappa^{2}\alpha_{\rm rad}\pm\sqrt{\kappa^{4}\alpha_{\rm rad}^{2}+12\kappa^{4}\alpha_{s}-4\kappa^{6}\alpha_{s}\vert\alpha_{s-A}\vert}}{-6+2\kappa^{2}\vert\alpha_{s-A}\vert}} \ .
\end{equation}
For the other values of $b$ one gets similar results, although the expressions are quite more cumbersome. We emphasize the fact that the presence of a minimum value of the scale factor in the early hot Big Bang implies the finiteness of geometrical quantities at the bounce, such as the Ricci curvature and torsion of the RC spacetime. For instance, in the ansatz $A_{\mu}=(\phi(t),0,0,0)$ one has
\begin{eqnarray}
T_{\alpha\beta\gamma}
&=&\kappa^{2}(s^{D}_{\alpha\beta\gamma}+2\lambda\kappa^{2}s^{D}_{\alpha[\beta\vert\rho}A_{\gamma]}A^{\rho}) \ .
\end{eqnarray}
Since $s^{D}_{\alpha\beta\gamma}$ is totally antisymmetric, then $K^{\lambda}=2T^{\lambda}=0$, so that using Eq. (\ref{Cterm}) we have
\begin{eqnarray}
R&=&\tilde{R}-\kappa^{4}\left[\lambda\kappa^{2}(2-\lambda\kappa^{2}\phi^{2})\phi^{2}\breve{s}^{2}-\dfrac{3}{2}\breve{s}^{2}\right]  \ .
\end{eqnarray}
In Case III, $\phi(a)\sim a^{4}$ and in general $s^{D}\sim\breve{s}\sim n(t)\sim a^{-3}$, therefore,
\begin{eqnarray}
R(a_{b})\sim\tilde{R}(a_{b})-2\alpha\lambda\kappa^{6}a_{b}^{2}+\beta\lambda^{2}\kappa^{8}a_{b}^{10}+\gamma a_{b}^{-6} \ ,
\end{eqnarray}
where $\alpha,\beta,\gamma$ are constants and $a_{b}$ is the scale factor at the bounce. Similarly the torsion components also remain finite. Let us note that the second and third terms in the expression above scale with $\sim a^{2}$ and  $\sim a^{10}$, respectively, which could imply a cosmological future singularity, occurring asymptotically when the scale factor goes to infinity.

\subsubsection{Early acceleration and cyclic cosmology}

\begin{figure*}[t!]
\centering
\includegraphics[width=7.0cm]{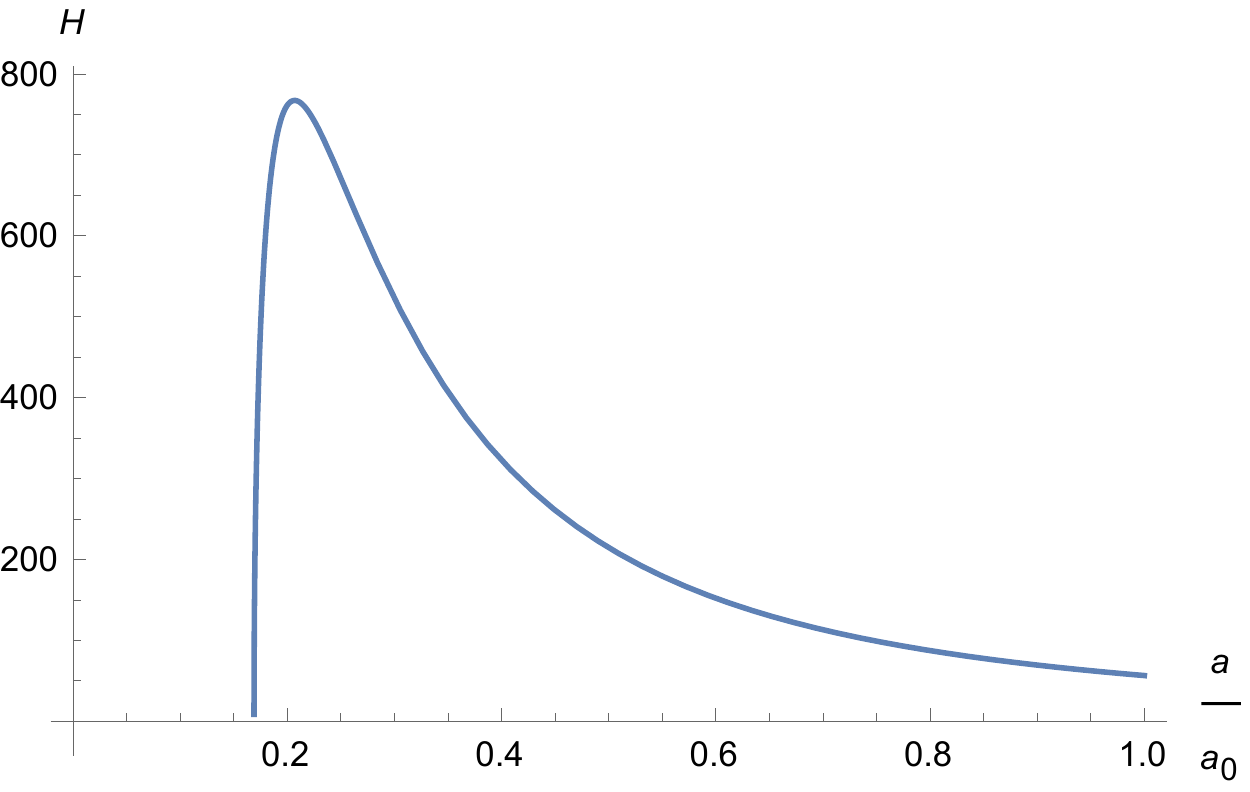}
   \hspace{1.5cm}
\includegraphics[width=7.0cm]{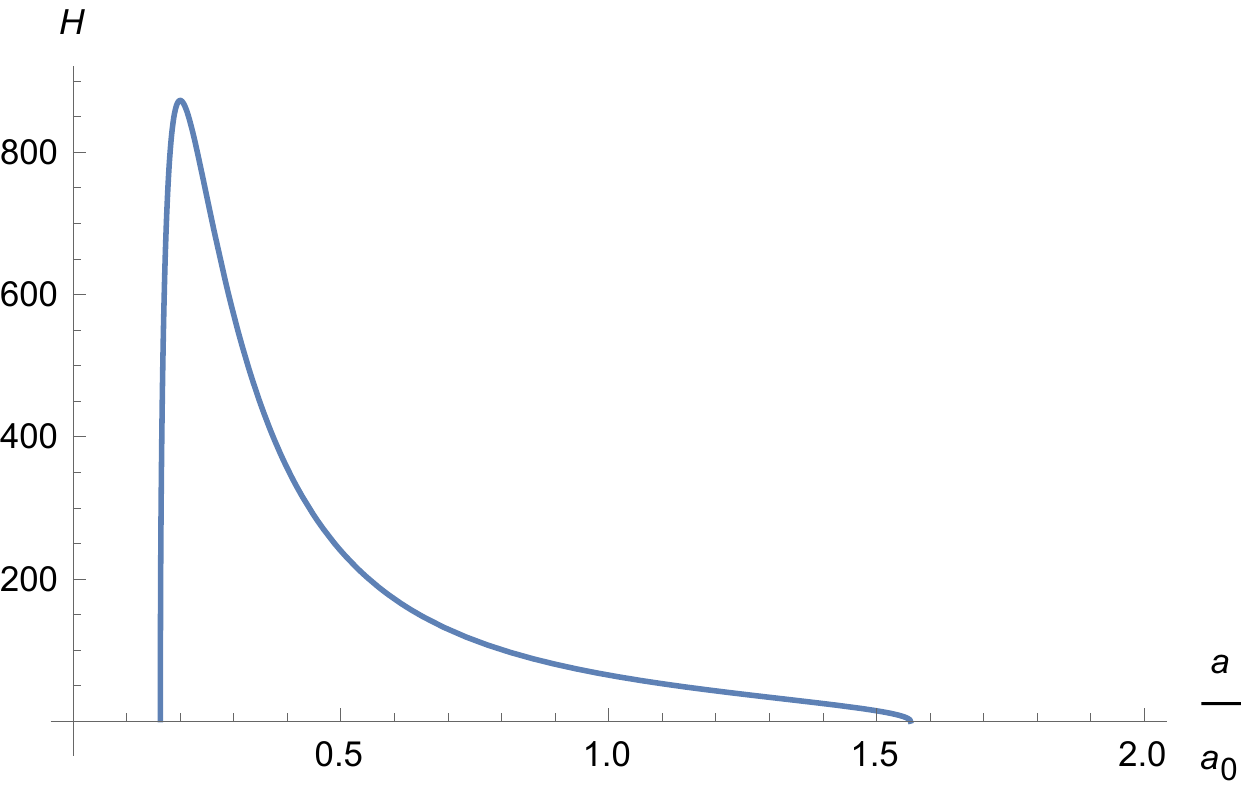}
\caption{Evolution of the Hubble parameter $H(a)$ with the scale factor $a/a_{0}$ for the ECDM model without (ECSK model, left) and with (right) the non-minimal couplings in the matter fields induced by torsion. These corrections to the effective energy density $\rho^{s-A}\sim\rho^{s}f(A)\sim a^{b}$ give raise to late-time effects, whereas  $\rho^{s}\sim -\kappa^{2}\breve{s}^{2}$ is the spin-spin interaction term that is responsible for the non-singular behaviour in the early Universe. The plot on the right shows a typical solution with a future bounce, a non-singular behaviour at the minimum of the scale factor, and a period of early accelerated expansion. All models we analysed, except case IV ($b=3$, $\rho^{s-A}>0$), show a typical cosmological behaviour as illustrated on the right plot, for the three spatial geometries $k=-1,0,1$.  The parameters used are: $\Omega_{r}=0.7$, $\Omega_{m}=0.32$, $\Omega_{s}=-0.02$, $H_{0}=68$, $\Omega_{k}=0.01$, $\alpha_{s-A}=-0.08$, $b=2$. }
\label{fig:1}
\end{figure*}

\begin{figure*}[t!]
\centering
\includegraphics[width=7.0cm]{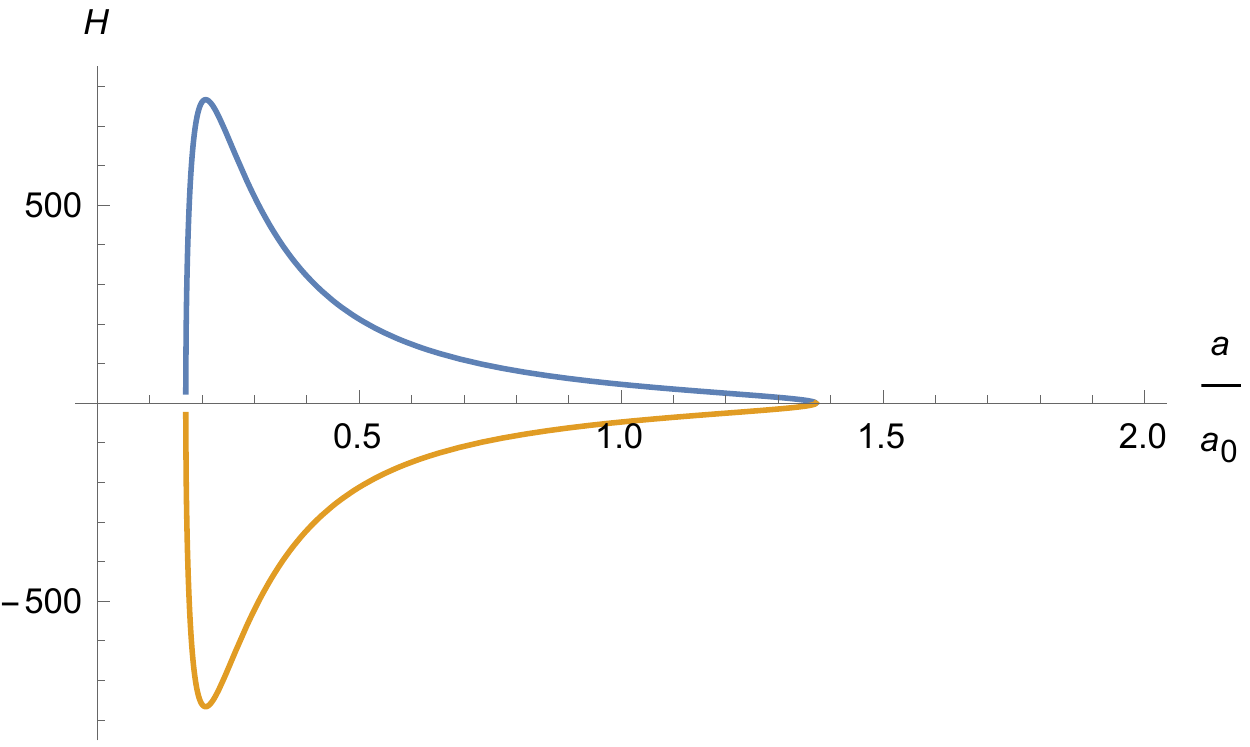}
   \hspace{1.5cm}
\includegraphics[width=7.0cm]{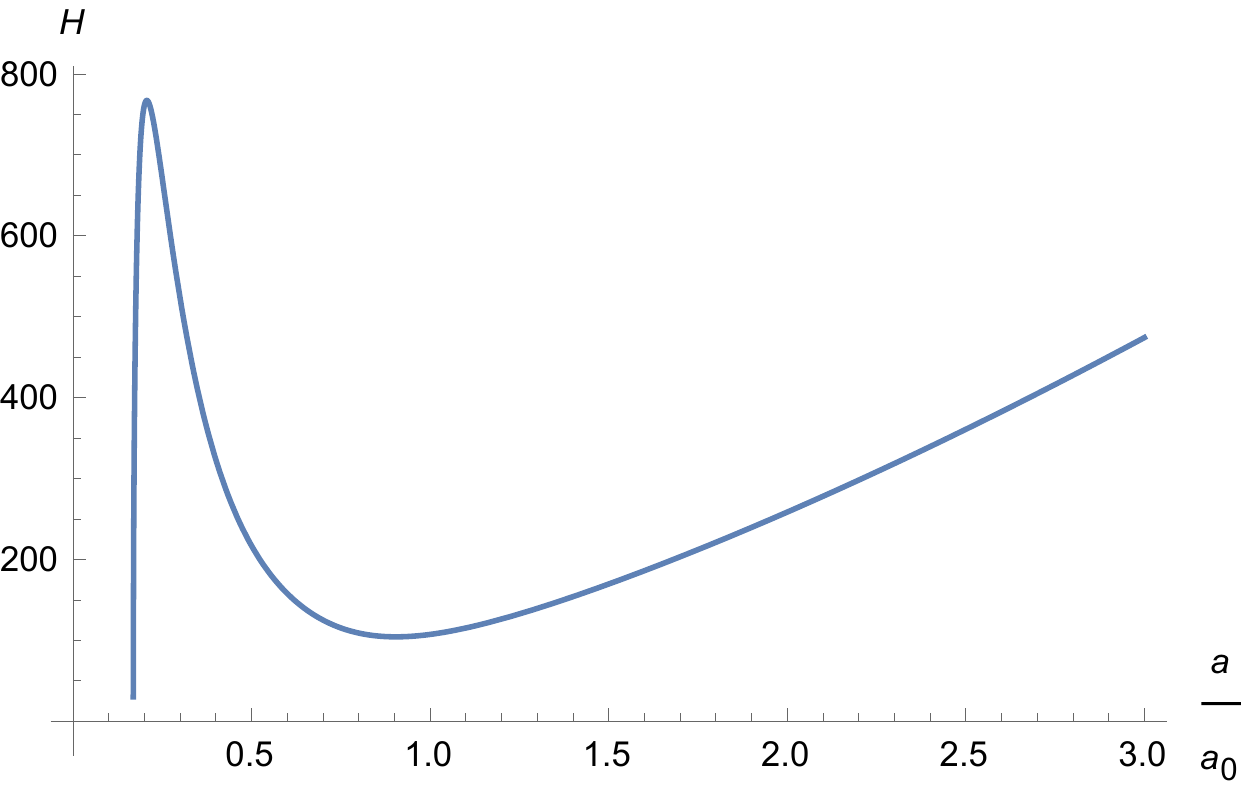}
\caption{On the left figure we can see the cyclic behaviour of ECDM model explicitly, with the two branches in $H(a)=\pm\sqrt{\rho^{\rm eff}(a)-k/a^{2}}$ smoothly joined together at the bounces. A period of early accelerated expansion is followed by decelerated expansion, bounce and accelerated contraction, decelerated contraction and again the bounce at the minimum of the scale factor, with the repetition of the cosmological cycle. On the right we have the relevant Case IV ($b=3$, $\rho^{s-A}>0$), where a late-time accelerated phase is also present.  The  parameters used are (left): $\Omega_{r}=0.7$, $\Omega_{m}=0$, $\Omega_{s}=-0.02$, $H_{0}=68$, $\Omega_{k}=-0.01$, $\alpha_{s-A}=-0.08$, $b=0$; (right): $\Omega_{r}=0.7$, $\Omega_{m}=0$, $\Omega_{s}=-0.02$, $H_{0}=68$, $\Omega_{k}=-0.01$, $\alpha_{s-A}=1.8$, $b=3$.}
\label{fig:2}
\end{figure*}

One can show that, for any $\lambda \neq 0$, in the cases studied above for $b=-2.88$, $b=2$, $b=3$ and $b=0$ (variation of Case IV), besides the minimum of the scale factor at the Big Bang there is a period of acceleration where the Hubble parameter increases until it reaches a maximum and starts decreasing (period of deceleration). This is valid for the spherical, flat, and hyperbolic spatial geometries. The effect of increasing the strength of the corrections to progressively higher values of $\lambda$ are different. For $b=-2.88$ (Case I) and for the three spatial geometries, both the value of the scale factor at the bounce and the ``instant'' of transition from positive acceleration towards deceleration tend to move into later times.
On the other hand, in the cases  $b=2$ (Case III) and $b=0$ (variation of Case IV) and also for $b=-2.88$, an increasing $\lambda$ reveals the relevance of a negative contribution to the energy density at later times. Indeed, for a critical value of such a contribution there will be a value of the scale factor for which the Hubble parameter vanishes (the deceleration and the expansion itself stops) and above that value it becomes imaginary, $H^{2}(a)<0$.

\begin{table*}[t!]
\centering
\begin{center}
\begin{tabular}{|c||c|c|c|c|}
\hline
        & Case I & Case III & Case IV & Case IV (var)  \\
        & $b=-2.88$ & $b=2$ & $b=3$ & $b=0$  \\
\hline
\hline
$A_{\mu}$      & $\phi\sim a^{0.78}$& $\phi\sim a^{4}$ & $A_{j}\sim a^{11/2}$ & $A_{j}\sim a^{4}$  \\
\hline
$\rho^{s-A}$ & $<0$ &  $<0$  &  $>0$  &  $<0$  \\
\hline
$w^{s-A}$   & $\approx -1/24$ & -5/3 & -2 & -1  \\
\hline
Early bounce ($a_{min}$)  & yes & yes & yes & yes  \\
\hline
Early acceleration & yes & yes & yes & yes  \\
\hline
Future bounce ($a_{max}$) & yes & yes & - & yes  \\
\hline
Late-time acceleration &  -& - & yes & -  \\
\hline
\end{tabular}
\caption{In this table one can see the main dynamical features of various cosmological scenarios studied in this paper. The cosmological dynamics is determined by the Friedman equations with spin-spin and non-minimal couplings effects (in the matter fields) induced by torsion. The late-time effects are dominated by the non-minimal interactions $\rho^{s-A}\sim a^{b}$.}
\label{tab1}
\end{center}
\end{table*}
\begin{table}[t!]
\centering
\begin{center}
\begin{tabular}{|c||c|c|}
\hline
        & Case I & Case III  \\
        & $b=-2.88$ & $b=2$  \\
\hline
\hline

Torsion & $T\sim\kappa^{2}s^{D}+\lambda\kappa^{4}s^{D}\phi^{2}$ & $T\sim\kappa^{2}s^{D}$ \\
         &  $\rightarrow 0$& $\rightarrow 0$  \\
\hline

$U(1)$ -   & $\mathcal{L}_{U1}\sim\lambda\kappa^{4}\breve{s}^{2}\phi^{2}$& $\mathcal{L}_{U1}\sim\lambda\kappa^{4}\breve{s}^{2}\phi^{2}$ \\
     Lagrangian    &  $\rightarrow 0$& $\sim a^{2}$ \\
\hline
\end{tabular}
\caption{In this table we illustrate that even though torsion is expected to decay, the $U(1)$-breaking Lagrangian does not necessarily decay too (see Case III above). Note that, as explained in the text, in the ansatz $A_{\mu}=(\phi,0,0,0)$, from Eq.(\ref{finalnewCartan}), one can see that $T\sim \kappa^{2}s^{D}+\lambda\kappa^{4}s^{D}\phi^{2}$ and while the first term always decays, the second  might not if $\phi\sim a^{m}$ with $m\geqslant3/2$. This is what happened in the alternative version of Case I, where from charge current arguments it was found that $\phi\sim a^{3}$, therefore implying a non-zero constant background torsion in homogeneous cosmologies.}
\label{tab}
\end{center}
\end{table}

The case of a constant energy density contribution ($b=0$) is particularly illuminating on this issue. From the Friedman equations (\ref{eq:F3}), the late-time cosmology of a positive constant energy density dominating asymptotically leads to the convergence of the Hubble parameter into a constant value of $H(a)$, but if the contribution from a negative energy density component starts to dominate, then the Hubble parameter is not well defined from the Friedman equations, as it becomes imaginary. This transition (when $H=0$) could be interpreted as a future bounce, and it is compatible with the idea of nature obeying, at least, the dominant energy condition $\rho\geqslant\vert p\vert$ (which implies the weak condition $\rho\geqslant 0$, $\rho+p\geqslant 0$), an interpretation that becomes quite clear in the flat case, $k=0$. Furthermore, due to the symmetry of the underlying Friedman equations this future bounce would be followed by a contraction, $H(a)<0$, gradually accelerated, then the contraction would move towards a decelerated contraction phase (since $H(a)$ has a local minimum)  until finally reaching the  minimum of the scale factor. At that point, the energy conditions and the requirement of a non-imaginary (real) Hubble parameter imply a non-singular behaviour and the new cycle of accelerated expansion followed by decelerated expansion would start.

This contracting behaviour is a natural path for the solution at the future bounce since there are two real solutions, $H(a)=\pm\sqrt{\rho^{\rm eff}(a)-k/a^{2}}$, corresponding to two branches of the possible cosmic history, in this case joined together at the two bounces. In both the early accelerated expansion (in branch 1) and in the sudden halt of the accelerated contraction (in branch 2) into a period of decelerated contraction, the effects due to the contribution of the dominant spin-spin (torsion induced) interaction will prevent a cosmic singularity. This cyclic behaviour is what happens in Cases I, III and IV (variation), as well as in several other models corresponding to different values of $b$. We summarize this discussion in Table \ref{tab1}, and depict these behaviours in Figs. \ref{fig:1} and \ref{fig:2}. We point out that this dynamics could be further explored by explicitly introducing a positive cosmological constant, though we shall deal in the next section with an effective cosmological constant out of the spin-spin interaction of fermionic vacuum condensates.

\subsection{Effective cosmological constant and dark-energy}

Let us now present three different results relevant for the cosmological constant/dynamical dark energy problem \cite{Ivanov:2016xjm} within ECDM theory. We begin by noting that one can easily show that if instead of $\overline{\breve{s}^{i}\breve{s}_{j}}\sim \overline{\breve{s}^{2}}\delta^{i}_{j}/3$ and $\overline{A^{i}A_{j}}\sim \overline{\vec{A}^{2}}\delta^{i}_{j}/3$ we take $\overline{\breve{s}^{i}\breve{s}_{j}}\sim \overline{\breve{s}^{2}}\delta^{i}_{j}$ and $\overline{A^{i}A_{j}}\sim \overline{\vec{A}^{2}}\delta^{i}_{j}$ , then Case IV corresponds to $w_{s-A}=-1$, and $A_{j}\sim a^{4}$, with $\rho_{s-A}\sim \rm constant$ ($b=0$). This yields an effective cosmological constant with an energy density scale set by $\lambda\kappa^{4}\hbar^{2}n^{2}_{\rm ref}A_{\rm ref}^{2}$, where $n_{\rm ref}$ is the fermion number density at some reference cosmic time. Indeed,  in this case we have
\begin{equation}
\rho^{\rm corr}\simeq-\kappa^{2}\breve{s}^{2}\left(\dfrac{3}{4}-\lambda\kappa^{2}\vec{A}^{2}\right) \ ,
\end{equation}
where
\begin{equation}
\rho^{s-A}=\rho^{\rm eff}_{\Lambda}=\dfrac{\lambda\kappa^{4}}{2}\beta_{s}n^{2}\vec{A}^{2}= {\rm const} \ ,
\end{equation}
with $\beta_{s}\sim\hbar^{2}$. As we saw previously, since $\vec{A}^{2}<0$, instead of having a positive cosmological constant effect and the resulting late-time acceleration one gets a future bounce with a transition from decelerated expansion into a period of accelerated contraction, in the cyclic scenario discussed above.

The second interesting solution corresponds to $b=3$ in the first version of Case IV. Here we have $\rho^{\rm s-A}\simeq-\kappa^{2}\breve{s}^{2}\dfrac{\lambda\kappa^{2}}{3}\vec{A}^{2}>0$ and
\begin{equation}
\rho^{\rm corr}\simeq -\alpha_{s}a^{-6}+\alpha_{s-A}a^{3}, \qquad \alpha_{s-A}>0 \ ,
\end{equation}
representing a non-singular cosmology with early acceleration (as in the other cases) but it also predicts a late-time accelerated expansion phase. This behaviour is driven by an effective dark energy effect supported by the term $\rho\sim a^{3}$ and arising from a non-minimal coupling in the matter fields induced by torsion, which starts dominating at later times.

The third result is motivated by the possibility of quark condensates in vacuum predicted by QCD, i.e., the effects of non-zero vacuum expectation values $\left< 0|\bar{\psi}\psi|0 \right>$. Indeed, in ECDM theory we can generalize the effective cosmological constant obtained in the literature of ECSK theory \cite{Poplawski:2010jv} as
\begin{eqnarray}
\rho^{\rm eff}_{\Lambda}&\sim & \dfrac{3\kappa^{2}}{4}\left<0\vert\breve{s}^{2}\vert 0 \right>+\lambda\kappa^{4}\Big[(\alpha+\zeta\lambda\kappa^{2}A^{2}) \left<0\vert\breve{s}^{2}\vert 0 \right>A^{2}
\nonumber \\
&&+(\beta+\varepsilon\lambda\kappa^{2}A^{2})\left<0\vert\breve{s}^{\mu}\breve{s}^{\nu}\vert 0 \right>A_{\mu}A_{\nu}\Big] \ ,
\end{eqnarray}
with $\alpha,\beta,\varsigma,\varepsilon$ constants, which depend on the above spin density vacuum expectation values and on the electromagnetic four-potential.  Since we are considering fermions, we will assume that these can form a condensate in vacuum and use the Shifman-Veinshtein-Zakharov vacuum state approximation, as in Ref. \cite{Poplawski:2010jv}. In such an approximation, the following expression is valid
 \begin{eqnarray}
\left< 0|\bar{\psi}\Gamma_{1}\psi\bar{\psi}\Gamma_{2}\psi|0 \right> &=&\dfrac{1}{12^{2}}\left(\text{tr}\,\Gamma_{1}\text{tr}\,\Gamma_{2}-\text{tr}(\Gamma_{1}\Gamma_{2})\right)
	\nonumber  \\
&&\times \left( \left< 0|\bar{\psi}\psi|0 \right> \right)^{2}\nonumber \ ,
 \end{eqnarray}
where $\Gamma_{1}$, $\Gamma_{2}$ are any matrix from the set $\lbrace I,\gamma^{i},\gamma^{[i}\gamma^{j]},\gamma^{5},\gamma^{5}\gamma^{i}\rbrace$. Then, for quarks, QCD predicts a non-zero expectation value of $\bar{\psi}\psi$ in vacuum
\begin{equation}
\left< 0|\bar{\psi}\psi|0 \right> \approx\lambda_{\rm QCD}^{3}\approx -(230\,{\rm MeV})^{3} \ ,
\end{equation}
in geometrical system of units. We then get the general result
\begin{equation}
\rho^{\rm eff}_{\Lambda}\sim (54\, {\rm meV})^{4}+f(A)\left(\left<0\vert\bar{\psi}\psi\vert0 \right>\right)^{2} \ ,
\end{equation}
where the second term is the modification in the prediction of the ECSK theory of fermions.

\subsubsection{Fermionic torsion}

From the expression $T_{\mu\nu}^{\rm eff}=T_{\mu\nu}+U_{\mu\nu}$, for the case of fermionic torsion, we have
\begin{equation}
T^{\Lambda}_{\mu\nu}=-\left[\dfrac{\kappa^{4}\lambda}{2}\left(A^{2}\breve{s}^{2}-(\breve{s}\cdot A)^{2}\right)+\dfrac{3}{4}\kappa^{2}\breve{s}^{\lambda}\breve{s}_{\lambda}\right] g_{\mu\nu} \ ,
\end{equation}
where we recall that $\breve{s}^{\mu}=\dfrac{\hbar}{2}\bar{\psi}\gamma^{\mu}\gamma^{5}\psi$. Therefore, we get an additional term contributing to an effective cosmological constant beyond the usual one coming from the spin-spin interaction already present in the ECSK model. We can then compute the expression for dark energy, in the ansatz $A_{\mu}=(0,\vec{A})$, as
\begin{eqnarray}
 \rho^{\rm eff}_{\Lambda}&=&-\dfrac{3\kappa^2}{4}\left< 0|\breve{s}_{j}\breve{s}^{j}|0\right> -\dfrac{\kappa^{4}\lambda}{2}\Big( \left< 0|\breve{s}_{j}\breve{s}^{j}|0 \right> A^{2}
 \nonumber \\
 &&
 -\left< 0|\breve{s}^{k}\breve{s}^{j}|0\right> A_{k}A_{j}\Big) \ ,
 \end{eqnarray}
 and after some algebra, we obtain
\begin{eqnarray}
\rho^{\rm eff}_{\Lambda}&\approx & -(54\,{\rm meV})^{4}  +\dfrac{\kappa^{4}\lambda\hbar^{2}}{3}
\left(\left< 0|\bar{\psi}\psi|0\right>\right)^{2}
\times
	\nonumber \\
&&\times  \left[\dfrac{2}{3}A^{2}
-\dfrac{1}{96}\left[(A_{1})^{2}+(A_{2})^{2}+(A_{3})^{2}\right]\right] \ .
\end{eqnarray}

In the ansatz $A_{\mu}=(\phi,0,0,0)$ we get instead
\begin{eqnarray}
 \rho^{\rm eff}_{\Lambda}&=&-\dfrac{3\kappa^2}{4}\left< 0|\breve{s}_{j}\breve{s}^{j}|0\right>  +\dfrac{\kappa^{4}\lambda}{2}\left( \left< 0|\breve{s}_{j}\breve{s}^{j}|0 \right> \phi^{2}\right) \ ,
 \end{eqnarray}
therefore
\begin{eqnarray}
\rho^{\rm eff}_{\Lambda}&\approx& -(54\,{\rm meV})^{4}-\kappa^{4}\lambda\hbar^{2}\dfrac{2}{9}\phi^{2}\times \left(\left< 0|\bar{\psi}\psi|0\right>\right)^{2} \ .
\end{eqnarray}
These expressions extend the results from the standard ECSK theory \cite{Poplawski:2010jv} by adding a dynamical dark energy term which depends on the four-potential ($\phi\sim a^{4}$ in Case III and $A_{j}\sim a^{11/2}$ in Case IV), during the $U(1)$-breaking symmetry phase induced by torsion. Let us point out that, as long as the minimal coupling between torsion and the bosonic four-potential takes place, the dynamical dark energy term is present. In other words, in the regimes in which the $U(1)$ breaking term in the bosonic Lagrangian (\ref{newMaxLagra}) is non-negligible the four-potential will evolve with the scale factor as it is explored in this work. Note that should $\lambda$ be considered as a scalar field then it would govern the transition for a (spontaneous) symmetry breaking regime, rather than having an explicit symmetry breaking as in the case where $\lambda$ is considered to be a constant coupling factor.

In absolute value, the result from the simple ECSK theory is much better than the $\sim 120$ order of magnitude discrepancy from observations (assuming GR with cosmological constant) with respect to the predictions from quantum field theory. In the ECDM model, and from the expressions above, in principle this result could be further improved depending on the ansatz taken for the four-potential.

\subsubsection{Full approach including the bosonic spin tensor}

Let us now consider the most general case in which torsion not only couples to the bosonic sector but it is also a result of the contribution from the total spin density including the spin density of bosons. Indeed, in such a case one has to consider Eqs. (\ref{newMaxLagra}) and (\ref{finalnewCartan}). Let us begin by isolating the following piece of the energy-momentum tensor (\ref{effectiveenergym})
\begin{eqnarray}
&&T^{{\rm eff}\Lambda}_{\mu\nu}=-\Bigg[\kappa^{2}\breve{s}^{2}\left(\dfrac{3}{4}+\dfrac{\lambda\kappa^{2}A^{2}}{2}\right)
+\dfrac{\lambda\kappa^{4}}{2} \times
\\
&&\quad \left[(2-\lambda\kappa^{2}A^{2})(A^{2}\breve{s}^{2}-(A\cdot\breve{s})^{2})
-(A\cdot\breve{s})^{2}\right]\Big]g_{\mu\nu} \ ,  \nonumber
\end{eqnarray}
which was derived in the ansatz $A_{\mu}=(\phi,0,0,0)$. So, we can write
\begin{equation}
T^{{\rm eff}\Lambda}_{\mu\nu}=-\Big[\kappa^{2}\breve{s}^{2}\left(\dfrac{3}{4}+\dfrac{\lambda\kappa^{2}\phi^{2}}{2}\right)
+\dfrac{\lambda\kappa^{4}}{2}(2-\lambda\kappa^{2}\phi^{2})\phi^{2}\breve{s}^{2} \Big]g_{\mu\nu},
\end{equation}
and therefore
\begin{eqnarray}
\rho^{\rm eff}_{\Lambda}\approx -(54\,{\rm meV})^{4}
-\dfrac{\kappa^{4}\lambda}{2}\phi^{2}\left[1+(2-\lambda\kappa^{2}\phi^{2})\right]
\left< 0|\breve{s}^{2}|0\right> , \nonumber
\end{eqnarray}
leading to
\begin{eqnarray}
\rho^{\rm eff}_{\Lambda}&\approx& -(54\,{\rm meV})^{4}
	\\
&&-\kappa^{4}\lambda\hbar^{2}\dfrac{2}{9}\phi^{2}\left[1+(2-\lambda\kappa^{2}\phi^{2})\right]
	\left(\left< 0|\bar{\psi}\psi|0\right>\right)^{2} \ . \nonumber
\end{eqnarray}
Moreover,  from Eq.(\ref{finalnewCartan}) we see that
$T\sim \kappa^{2}s^{D}+\lambda\kappa^{4}s^{D}\phi^{2}$, (since $\tilde{s}=0$). While the first term always decays, the second  might decay or not (if $\phi\sim a^{m}$ with $m\geqslant3/2$). However, the predicted behaviour for $\phi$ rests on the validity of the extended Maxwell Lagrangian. As one can see in Eq.(\ref{newMaxLagra}) the first term in the $U(1)$-breaking term scales as $\lambda T^{2}\phi^{2}$, and therefore if $m\geqslant 3$ it does not decay.

In Case I we obtained the approximate solution $\phi\sim a^{0.78}$, so that the dark energy effect above is valid only during the transient $U(1)$ broken phase since in this case the $U(1)$-breaking term in Eq.(\ref{newMaxLagra}), $\lambda T^{2}\phi^{2}$, decays with the increasing scale factor (see Table II). Note that the $\phi\sim a^{0.78}$ behaviour was deduced from a simplified and not very robust approximation and, as we shall show below, the generalized charge conservation equation seems to suggest that $\phi\sim a^{3}$ also in this case. If so, then interestingly the torsion tensor $T\sim \kappa^{2}s^{D}+\lambda\kappa^{4}s^{D}\phi^{2}$ does not decay to zero, leaving a constant torsion background. Moreover, as can be seen in Eq.(\ref{newMaxLagra}) the first term in the $U(1)$-breaking term $\lambda T^{2}\phi^{2}$ also remains constant.

\subsection{Coupling to Maxwell dynamics}

By varying Eq.(\ref{eq:Maxfull}), together with the minimal coupling term $j^{\alpha}A_{\alpha}$, with respect to $A_{\mu}$, one gets
\begin{equation}
\nabla_{\mu}F^{\mu\nu}=\lambda^{-1}j^{\nu} \ ,
\end{equation}
where $j^{\nu}=q\bar{\psi}\gamma^{\nu}\psi$ is the Dirac charge current four-vector. This equation can be conveniently rewritten as
\begin{equation} \label{eq:ECDEem}
\qquad \tilde{\nabla}_{\mu}\tilde{F}^{\mu\nu}=\lambda^{-1}(j^{\nu}+J^{\nu}) \ ,
\end{equation}
where we have defined the torsion-induced four-current
\begin{eqnarray}
\label{newcurrentgeneral}
J^{\nu}&=&-\lambda\Big[2(K^{\nu}_{\;\;\lambda\mu}K^{\gamma[\mu\lambda]}+K_{\lambda}K^{\gamma[\lambda\nu]})A_{\gamma}
+K^{\nu}_{\;\;\lambda\mu}\tilde{F}^{\mu\lambda}
	\nonumber \\
&&\qquad +K_{\lambda}\tilde{F}^{\lambda\nu}
+2\tilde{\nabla}_{\mu}\left(K^{\gamma[\mu\nu]}A_{\gamma}\right)\Big]  \ ,
\end{eqnarray}
with $K_{\lambda}\equiv K^{\alpha}_{\;\;\lambda\alpha}$.\footnote{As can be seen in the expression for the Lagrangian in Eq.(\ref{newMaxLagra}), or in the field equations  above, the terms quadratic in the contortion or, equivalently, in the spin density, resemble Proca-like terms. From this analogy, the coupling between the electromagnetic four-potential and the spacetime torsion provides an effective mass for the photon $m_{\gamma}^{2}\sim \lambda T^{2}$ in physical conditions where torsion is non-negligible and the $U(1)$-breaking phase transition takes place. The generalized current can also be written as
\begin{eqnarray}
\label{4current}
J^{\nu}&=&-\lambda\Big[2(T^{\nu}_{\;\;\lambda\mu}T^{\gamma\mu\lambda}+2T_{\lambda}T^{\gamma\lambda\nu})A_{\gamma}\,
\nonumber \\
&&+T^{\nu}_{\;\;\lambda\mu}\tilde{F}^{\mu\lambda}+2T_{\lambda}\tilde{F}^{\lambda\nu}
+2\tilde{\nabla}_{\mu}(T^{\gamma\mu\nu}A_{\gamma})\Big] \ ,
\end{eqnarray}
where we have used the fact that contortion is antisymmetric in the first two indices and also that $K^{\nu}_{\;\;[\lambda\mu]}=T^{\nu}_{\;\;\lambda\mu}$ and $K_{\lambda}=2T_{\lambda}$.}

On the other hand, the generalized conservation equation can be written as
\begin{equation}
\label{chargeequation}
\tilde{\nabla}_{\nu}j^{\nu}=-\tilde{\nabla}_{\nu}J^{\nu} \ ,
\end{equation}
or, alternatively, as
\begin{equation}
\nabla_{\nu}j^{\nu}=\dfrac{\lambda}{2}\left[ \nabla_{\nu},\nabla_{\mu}\right] F^{\mu\nu} \ ,
\end{equation}
where
\begin{equation}
\left[\nabla_{\nu},\nabla_{\mu}\right] F^{\mu\nu}=R^{\mu}_{\;\;\varepsilon\nu\mu}F^{\varepsilon\nu}+R^{\nu}_{\;\;\varepsilon\nu\mu}F^{\mu\varepsilon}+2T^{\gamma}_{\;\;\nu\mu}\nabla_{\gamma}F^{\mu\nu} \ ,
\end{equation}
is the commutator of covariant derivation of an antisymmetric $(0,2)$-tensor in RC spacetime. This expression is valid for the RC spacetime geometry, and the only requirement is a Maxwell-like bosonic field minimally coupled to the RC geometry.

The induced four-current correction term $J^{\nu}$ is due to the presence of non-minimal couplings between $A_{\mu}$ and the spinors $\psi,\bar{\psi}$ and bosonic self-interactions, both effects induced by torsion. It can be obtained by substituting the Cartan equations in (\ref{4current}), or by direct variation of the effective Maxwell Lagrangian (\ref{BosonLagcorr}). This torsion-induced current $J^{\nu}$ is given by
\begin{eqnarray}
J^{\nu}&=&\lambda\kappa^{2}\Big[\tilde{F}_{\alpha\beta}\left(\lambda A^{[\alpha}\tilde{F}^{\beta]\nu}+2A^{[\alpha}\tilde{s}^{\beta]}A^{\nu}X(A)\right)\nonumber \\
&&+2\tilde{F}^{\nu}_{\;\;\beta}\left({F}^{\beta}_{\;\;\lambda}A^{\lambda} +2{s}^{\beta}Y(A)\right)\nonumber \\
&&+\lambda^{2}\kappa^{2}\left(A^{\nu}\tilde{F}^{\alpha\lambda}A_{\lambda}+A^{2}\tilde{F}^{\alpha\nu}\right)\tilde{F}_{\alpha\gamma}A^{\gamma}\nonumber \\
&&+(A^{\nu}\tilde{s}^{2}-2\tilde{s}^{\nu}(A\cdot\tilde{s}))Z(A) \\
&&+A^{\nu}(A^{2}\tilde{s}^{2}-(A\cdot\tilde{s})^{2})W(A)\nonumber \\
&&-\kappa^{2}(A^{\nu}\breve{s}^{2}-\breve{s}^{\nu}(A\cdot\breve{s}))\Big]
-\tilde{\nabla}_{\mu}\left(\dfrac{\partial\mathcal{L}_{\rm corr}^{\rm M}}{\partial (\tilde{\nabla}_{\mu}A_{\nu})} \right),\nonumber
\end{eqnarray}
where the last term is computed as
\begin{eqnarray}
&&\dfrac{\partial\mathcal{L}_{\rm corr}^{\rm M}}{\partial (\tilde{\nabla}_{\mu}A_{\nu})}=2\lambda^{2}\kappa^{2}\Big(A^{[\mu}\tilde{F}^{\nu]\lambda}A_{\lambda}+\tilde{F}^{\alpha[\mu}A^{\nu]}A_{\alpha}
	\nonumber \\
&&
\qquad -\tilde{F}^{[\mu}_{\;\;\beta}A^{\nu]}A^{\beta}\Big)
+4\lambda\kappa^{2}A^{[\mu}\tilde{s}^{\nu]}\dfrac{1-\lambda\kappa^{2}A^{2}}{2+\lambda\kappa^{2}A^{2}}
 \\
&&\qquad +\lambda^{3}\kappa^{4}A^{2}\tilde{F}^{[\mu}_{\;\;\gamma}A^{\nu]}A^{\gamma} \nonumber   \ ,
\end{eqnarray}
and we have introduced the definitions
\begin{eqnarray}
X(A)&\equiv & -\dfrac{6\lambda\kappa^{2}}{(2+\lambda\kappa^{2}A^{2})^{2}} \,,
  \qquad
\nonumber \\
Y(A)&\equiv& -\dfrac{1-\lambda\kappa^{2}A^{2}}{2+\lambda\kappa^{2}A^{2}} \,,
	\nonumber \\
Z(A)&\equiv & \dfrac{2\kappa^{2}(1-(2-\lambda\kappa^{2}A^{2}))}{2+\lambda\kappa^{2}A^{2}} \,,
	\nonumber \\
W(A)&\equiv &\Big[4\lambda\kappa^{2}\Big((2+\lambda\kappa^{2}A^{2})(\lambda\kappa^{2}A^{2}-1) \,,
	\nonumber \\
&&-\left(1-\lambda\kappa^{2}A^{2}(2-\lambda\kappa^{2}A^{2})\right)\Big)\Big]/(2+\lambda\kappa^{2}A^{2})^{3} \nonumber \ .
\end{eqnarray}
These highly involved expressions can be interpreted as non-linear electrodynamics with non-minimal couplings between fermionic matter (spinors) and electromagnetic fields induced by the RC spacetime geometry.

\subsubsection{Maxwell fields in fermionic background torsion}

In the case of fermionic torsion (neglecting the contribution from the spin tensor of the bosonic field), the bosonic Lagrangian is simplified to (\ref{BosonLagsimple}). Under the assumption of the random spin distribution we obtain
\begin{eqnarray}
\label{torsioninducedcurrent}
J^{\nu}&=&-\kappa^{4}\lambda \left(\breve{s}^{2}A^{\nu}-(\breve{s}\cdot A)\breve{s}^{\nu}\right) \ .
\end{eqnarray}
Since we take $\breve{s}^{\lambda}$ to be spatial, we then have $J^{0}=-\kappa^{4}\lambda\breve{s}^{2}\phi$ and $J^{i}=0$ for $A_{\mu}=(\phi,0,0,0)$, while $J^{0}=0$ and $J^{i}=-\kappa^{4}\lambda \left(\breve{s}^{2}A^{i}-(\breve{s}\cdot A)\breve{s}^{i}\right)$ for $A_{\mu}=(0,\vec{A})$. In the last expression, using the previous assumptions after an average procedure, i.e., $\breve{s}^{i}\breve{s}_{j}=\breve{s}^{2}\delta^{i}_{j}/3$, we obtain  $J^{i}=-\dfrac{2}{3}\kappa^{4}\lambda\breve{s}^{2}A^{i}$.

\subsubsection{Full approach, including the bosonic spin tensor}

In this case we will again consider matter with a random distribution of fermionic spins, where we neglect all quantities linear in the Dirac spin, leaving only the quadratic ones which do not vanish after macroscopic averaging. Taking the ansatz $A_{\mu}=(\phi,0,0,0)$ we find
\begin{eqnarray}
J^{\nu}&=&-\lambda\kappa^{4}
\left[A^{\nu}\breve{s}^{2}-\breve{s}^{\nu}(A\cdot\breve{s})\right] \ ,
\end{eqnarray}
with $J^{0}=-\kappa^{4}\lambda\breve{s}^{2}\phi$ and $J^{i}=0$, just as we had in the case of a background fermionic torsion. Then, using the conservation equation (\ref{chargeequation}) we are led to $J^{0}\sim a^{-3}$, with $\phi\sim a^{3}$. Since this seems to be a more robust result than the $\phi\sim a^{0.78}$ previously used, if we go back to the fluid description in Case I, we then get $\rho^{s-A}\sim B+Ca^{6}$ with $B$ and $C$ negative constants. This fluid component manifests its effects in the evolution of the Hubble rate at late times implying an anticipation of the future bounce into earlier times, in comparison with the other cosmological solutions with future bounce.

On the other hand, taking into account the ansatz $A_\mu=(0,\vec{A})$, Maxwell's equations can be written as
\begin{equation}
\ddot{A}_{i}+H\dot{A}_{i}=\lambda^{-1}(j^{i}+J^{i}) \ ,
\end{equation}
and we can take $j^{i}\simeq 0$, on average. Alternatively, we have
\begin{equation}
\label{MaxeqA}
A''_{i}+A'_{i}\left(\dfrac{1}{a}+\dfrac{H'}{H}\right)+\dfrac{3}{a^{2}H}A_{i} =\lambda^{-1}\dfrac{J_{i}}{a^{2}H^{2}} \ ,
\end{equation}
where $H=H(a)$ and here the prime denotes a derivative with respect to the scale factor. In this case, this equation together with the Friedman equation
\begin{equation}
H^{2}(a)=\frac{\kappa^{2}}{3}\left(\alpha_{\rm rad}a^{-4}-\kappa^{2}\alpha_{s}a^{-6}+\rho^{s-A}(a)\right)-ka^{-2} \ ,
\end{equation}
determine the dynamics for the relevant degrees of freedom in the early Universe.

\subsection{Generalized Hehl-Datta (Dirac) equation in a cosmological context and matter/anti-matter asymmetry.} \label{secIVB}

\subsubsection{Fermionic torsion}

 The full cosmological dynamics is contained in the Friedman equations (\ref{eq:F1}) and (\ref{eq:F2}), the equation for the four-potential (\ref{MaxeqA}), and the Dirac equation in a FLRW background.
To derive such dynamics consider first the Dirac action in a RC spacetime given by the Lagrangian density in Eq.(\ref{DiracLag}), for the case of fermionic torsion (\ref{fermionictorsion}). This yields the Fock-Ivanenko-Heisenberg-Hehl-Datta equation \cite{Hehl-Data}
\begin{equation}
\label{eq:HehlData}
i\hbar \gamma^{\mu}\tilde{D}_{\mu}\psi-m\psi =
\dfrac{3\kappa^2\hbar^{2}}{8} (\bar{\psi}\gamma^{\nu}\gamma^{5}\psi)\gamma_{\nu}\gamma^{5}\psi \ .
\end{equation}
For cosmological applications it is useful to consider the comoving time variable $d\eta=dt/a(\eta)$, and the FLRW metric in its conformally flat expression\footnote{Note that, as explicitly shown in \cite{Iihoshi:2007uz,Gron1,Gron2,Gron3}, it is possible to find a system of coordinates where this formula is valid even for the open ($k=-1$) and closed ($k=1$) FLRW scenarios, since the Weyl tensor vanishes in all these cases.}
\begin{equation}
g_{\mu\nu}=a^{2}(\eta)\eta_{\mu\nu} \ .
\end{equation}
Then, we can use the identity
\begin{equation}
\gamma^{\mu}\tilde{D}_{\mu}\psi=a^{-\frac{5}{2}}(\eta)\gamma^{b}\partial_{b}\left( a^{\frac{3}{2}}(\eta)\psi\right) \ ,
\end{equation}
with $b=0,1,2,3$, to arrive at the Hehl-Datta (Dirac) equation in a FLRW background
\begin{eqnarray}
i\hbar\gamma^{0}\chi'=ma\chi+\dfrac{3\kappa^2\hbar^{2}}{8}a^{-2}(\bar{\chi}\gamma^{\nu}\gamma^{5}\chi)\gamma_{\nu}\gamma^{5}\chi,\nonumber
\end{eqnarray}
where
\begin{equation}
\chi(\eta)\equiv a^{\frac{3}{2}}(\eta)\psi \,, \qquad \bar{\chi}(\eta)\equiv a^{\frac{3}{2}}(\eta)\bar{\psi}\,,
\end{equation}
and the derivative is now performed with respect to the conformal time $\eta$.

Analogously, the generalized Hehl-Datta (Dirac) equation, including the non-minimal interaction with the electromagnetic four-potential in the case of fermionic torsion, can be easily derived from equations (\ref{newDiracLagsimple}) and (\ref{BosonLagsimple}) and is given by\footnote{Note, however, that if one performs the variational principle from $\mathcal{L}_{m}$ without substituting the torsion tensor via Cartan relations (\ref{fermionictorsion}) and only make such a replacement after the derivation of the dynamical equations, then in this case of fermionic torsion one arrives again at the usual Hehl-Datta equation.}
\begin{eqnarray}
 i\hbar \gamma^{\mu}\tilde{D}_{\mu}\psi+\left(q\gamma^{\mu}A_{\mu}-\dfrac{\kappa^2\lambda\hbar}{4}f^{\rho}\gamma_{\rho}\gamma^{5}-m\right)\psi
	\nonumber \\
 = \left(\dfrac{\kappa^{4}\lambda\hbar^{2}}{2}A^{2}+\dfrac{3\kappa^2\hbar^{2}}{8}\right) (\bar{\psi}\gamma^{\nu}\gamma^{5}\psi)\gamma_{\nu}\gamma^{5}\psi
	\nonumber	\\
 \qquad  -\dfrac{\kappa^{4}\lambda\hbar^{2}}{2}(\bar{\psi}\gamma^{\beta}\gamma^5\psi)\gamma_{\lambda}\gamma^5\psi A_{\beta}A^{\lambda} \ ,
\end{eqnarray}
and in the background of a FLRW cosmological metric it becomes
\begin{eqnarray}
&& i\hbar\gamma^{0}\chi'+\left(q\gamma^{\mu}A_{\mu}-\dfrac{\kappa^2\lambda\hbar}{4}f^{\rho}\gamma_{\rho}\gamma^{5}-m\right)a\chi
	\nonumber \\
&& \qquad  =\left(\dfrac{\kappa^{4}\lambda\hbar^{2}}{2}A^{2}+\dfrac{3\kappa^2\hbar^{2}}{8}\right)a^{-2} (\bar{\chi}\gamma^{\nu}\gamma^{5}\chi)\gamma_{\nu}\gamma^{5}\chi
	 \nonumber	 \\
&& \qquad \qquad -\dfrac{\kappa^{4}\lambda\hbar^{2}}{2}a^{-2}(\bar{\chi}\gamma^{\beta}\gamma^5\chi)\gamma_{\lambda}\gamma^5\chi A_{\beta}A^{\lambda} \ ,
\end{eqnarray}
with a similar dynamical (diffusion-like) cubic equation for $\bar{\chi}$. In these equations, $\gamma_{\mu}=e^{b}_{\;\mu}\gamma_{b}$ and  $\gamma^{\nu}=e_{c}^{\;\nu}\gamma^{c}$, where the tetrads in our coordinates become $e^{b}_{\;\mu}=\delta^{b}_{\;\mu}a$ and $e_{c}^{\;\nu}=\delta_{c}^{\;\nu}a^{-1}$, which follows from $e^{a}_{\;\alpha}e^{b}_{\;\beta}\eta_{ab}=g_{\alpha\beta}$ $(a,b,c=0,1,2,3)$ and its inverse relation. We have also assumed homogeneous fields, so that each variable depends only on the conformal time. Accordingly, $f^{\nu}$ is given by Eq.(\ref{smallf}), where the only non-vanishing components of the Faraday tensor in this system of coordinates are $\tilde{F}_{0j}(\eta)=\partial_{\eta}A_{j}=a(\eta)\dot{A}_{j}$.

In the ansatz of $A_{\mu}=(\phi(t),0,0,0)$ we get
\begin{eqnarray}
i\hbar\gamma^{0}\chi'+\left(q\gamma^{0}\phi-m\right)a\chi =
	\nonumber \\
\Bigg(\dfrac{\kappa^{4}\lambda\hbar^{2}}{2}\phi^{2}+\dfrac{3\kappa^2\hbar^{2}}{8}\Bigg)a^{-2} (\bar{\chi}\gamma^{\nu}\gamma^{5}\chi)\gamma_{\nu}\gamma^{5}\chi
	\nonumber \\
-\dfrac{\kappa^{4}\lambda\hbar^{2}}{2}a^{-2}\phi^{2}(\bar{\chi}\gamma^{0}\gamma^5\chi)\gamma_{0}\gamma^5\chi \ ,
\end{eqnarray}
with $\phi\sim a^{4}$ (Case III), yields
\begin{eqnarray}
&&i\hbar\gamma^{0}\chi'+\left(q\gamma^{0}Ca^{4}-m\right)a\chi = \\
&&\dfrac{3\kappa^2\hbar^{2}}{8}a^{-2} (\bar{\chi}\gamma^{\nu}\gamma^{5}\chi)\gamma_{\nu}\gamma^{5}\chi
\dfrac{\kappa^{4}\lambda\hbar^{2}}{2}Ca^{6}(\bar{\chi}\gamma^{k}\gamma^5\chi)\gamma_{k}\gamma^5\chi \ , \nonumber
\end{eqnarray}
where $C$ is an integration constant.

\subsubsection{Full approach including the bosonic spin tensor}

To consider the general case, i.e, taking into account the bosonic contribution to the spin tensor and therefore to torsion, we start from the general expression of the Dirac equation minimally coupled to the RC geometry
\begin{equation}
\label{DiracRC}
i\hbar \gamma^{\mu}\tilde{D}_{\mu}\psi+\left(q\gamma^{\mu}A_{\mu}-m\right)\psi =-\dfrac{3\hbar}{2}\breve{T^{\lambda}}\gamma_{\lambda}\gamma^5\psi.
\end{equation}
We now simply substitute the axial torsion vector in (\ref{axialtorsion}), derived from the full Cartan equations (\ref{finalnewCartan}).
After some algebra, we obtain the following extended Dirac (cubic) equation\footnote{
If we consider instead the total matter Lagrangian
\begin{equation}
{\cal L}_{m}={\cal L}_{\rm D}+{\cal L}_{\rm M}+j^{\mu}A_{\mu} \ ,
\end{equation}
with $L_{\rm D}$ given by (\ref{newDiracLag}) and ${\cal L}_{\rm M}=\tilde{\cal L}_{\rm M}+{\cal L}^{\rm M}_{\rm corr}$ with ${\cal L}^{\rm M}_{\rm corr}$ as in Eq.(\ref{BosonLagcorr}), i.e., if we substitute Cartan's equations at the Lagrangian level and then vary with respect to spinors, we arrive at a similar Dirac equation with more complicated functions of $A$ and $\tilde{F}$.}
\begin{eqnarray}
\label{newHehlData}
&&i\hbar \gamma^{\mu}\tilde{D}_{\mu}\psi+\left(q\gamma^{\mu}A_{\mu}-m\right)\psi =f(A)(\bar{\psi}\gamma^{\nu}\gamma^{5}\psi)\gamma_{\nu}\gamma^{5}\psi\nonumber \\
&&\quad +\alpha^{\lambda}_{\;\alpha}(A)(\bar{\psi}\gamma^{\alpha}\gamma^5\psi)\gamma_{\lambda}\gamma^5\psi +\beta^{\lambda}(A,\tilde{F})\gamma_{\lambda}\gamma^5\psi \ ,
\end{eqnarray}
where we have defined
\begin{eqnarray}
f(A) &\equiv & \dfrac{3\kappa^{2}\hbar^{2}}{8}+\dfrac{\lambda\kappa^{4}\hbar^{2}}{4}A^{2},
	\nonumber \\
\alpha^{\sigma\varepsilon}(A) &\equiv & -\lambda\kappa^{4}\hbar^{2}A^{\sigma}A^{\varepsilon},
	\nonumber \\
\beta^{\lambda}(A,\tilde{F})&\equiv &- \dfrac{\lambda\kappa^{2}\hbar}{2}\epsilon^{\lambda\alpha\beta\gamma}A_{[\alpha}\tilde{F}_{\beta\gamma]} \ .
\end{eqnarray}
Therefore, in the context of FRLW cosmology
\begin{eqnarray}
 i\hbar\gamma^{0}\chi' &+&\left[q\gamma^{\mu}A_{\mu}-\beta^{\rho}(A,\tilde{F})\gamma_{\rho}\gamma^{5}-m\right] a\chi
	\nonumber \\
&=& f(A)a^{-2} (\bar{\chi}\gamma^{\nu}\gamma^{5}\chi)\gamma_{\nu}\gamma^{5}\chi
	\nonumber \\
&& \quad +\alpha^{\beta\lambda}(A)a^{-2}(\bar{\chi}\gamma_{\beta}\gamma^5\chi)\gamma_{\lambda}\gamma^5\chi \ ,
\end{eqnarray}
and in the ansatz $A_{\mu}=(\phi(t),0,0,0)$ we have
\begin{eqnarray}
f(\phi) &\equiv & \dfrac{3\kappa^{2}\hbar^{2}}{8}+\dfrac{\lambda\kappa^{4}\hbar^{2}}{2}\phi^{2}\,,
	\nonumber \\
\alpha^{00}(\phi) &=& -\dfrac{\lambda\hbar^{2}\kappa^{4}}{4}\phi^{2}\,,
	\nonumber \\
\beta^{\alpha}&=&0 \ ,
\end{eqnarray}
which yields the result
\begin{eqnarray}
i\hbar\gamma^{0}\chi'&+&\left(q\gamma^{0}\phi-m\right)a\chi =
f(\phi)a^{-2} (\bar{\chi}\gamma^{\nu}\gamma^{5}\chi)\gamma_{\nu}\gamma^{5}\chi
	\nonumber \\
&&+\alpha^{00}(\phi)a^{-2}(\bar{\chi}\gamma_{0}\gamma^5\chi)\gamma_{0}\gamma^5\chi \ .
\end{eqnarray}

Using the result derived from the generalized charge conservation, $\phi(a)\sim a^{3}$, we then get $f(\phi)\sim {\rm const}+a^{6}$, and $\alpha^{00}\sim -a^{6}$.
The equation above is coupled to the equation for the adjoint spinors
\begin{eqnarray}
i\hbar\bar{\chi}'\gamma^{0}&-&a\bar{\chi}\left(q\gamma^{0}\phi-m\right)=
-f(\phi)a^{-2} (\bar{\chi}\gamma^{\nu}\gamma^{5}\chi)\gamma_{\nu}\bar{\chi}\gamma^{5}
	\nonumber \\
&&-\alpha^{00}(\phi)a^{-2}(\bar{\chi}\gamma_{0}\gamma^5\chi)\gamma_{0}\bar{\chi}\gamma^{5} \ .
\end{eqnarray}
Under a charge conjugation (C) operation $\psi\rightarrow -i\gamma^{2}\psi^{*}\equiv \psi^{ch}$, corresponding to the Dirac equation for antiparticles, we have instead
\begin{eqnarray}
&& i\hbar\gamma^{0}(\chi^{ch})'-\left(q\gamma^{0}\phi+m\right)a\chi^{ch}
	\nonumber \\
&& \qquad  =-f(\phi)a^{-2} (\bar{\chi^{ch}}\gamma^{\nu}\gamma^{5}\chi^{ch})\gamma_{\nu}\gamma^{5}\chi^{ch}
	\nonumber \\
  &&\qquad \quad-\alpha^{00}(\phi)a^{-2}(\bar{\chi^{ch}}\gamma_{0}\gamma^5\chi^{ch})\gamma_{0}\gamma^5\chi^{ch} \ .
\end{eqnarray}

Since the dynamics for (homogeneous) spinors representing fermions and anti-fermions are different (the cubic terms have changed signs) and are therefore related to different decay laws, this is highly relevant for the topic of matter/anti-matter asymmetry in the early Universe.  To illustrate this idea qualitatively one could simply consider two different orbits in the space ($y'$,$y$) for different values of $\eta$ in the following dynamical scenario $y'(y;\eta)=y\left[B\eta^{1/2}\pm (C\eta^{2}+D\eta^{-1})y^{2}\right]$, which is motivated from the above equations. Such a simplified but quite general behaviour can be obtained by considering $a\sim t^{2/(3+3w^{\rm dom})}$ and $w^{\rm dom}=1$ for the dominant fluid in the early Universe, leading to $\rho^{\rm dom}\sim a^{-6}$, $a\sim t^{1/3}$, $\eta\sim t^{2/3}$, $t\sim\eta^{3/2}$, and therefore $a\sim\eta^{1/2}$. Of course in our model things are more complicated since we have four component spinors, but the trajectories associated to the $+$ and $-$ sign above (corresponding to fermions and anti-fermions, respectively) already illustrate how a matter/anti-matter asymmetry could be generated in the torsion era of the early Universe. Although there are no parity-breaking terms in our model (which is one of the Zakharov requisites, together with $C$ breaking, for a successful mechanism generating matter/anti-matter asymmetry), our model does include  an explicit $C$-symmetry breaking. One could go beyond the minimal coupling of fermions and torsion to include such parity-breaking terms, as these appear naturally in some quadratic models of Poincar\'{e} gauge theory of gravity.

It has been shown (see \cite{Poplawski:2011xf} and  references therein)  that by solving the Dirac-Hehl-Datta equation, in the approximation of zero curvature and constant background torsion, the energy levels are different for fermions and anti-fermions. This can have consequences for the matter/anti-matter asymmetries in the context of baryogenesis. Another interesting consequence is the fact that, depending on whether a fermion has its spin aligned or anti-aligned with the background spin (torsion) density field, its energy is different and transitions between these levels can produce emission/absorption lines with a kind of hyperfine structure. This is reminiscent of the Zeeman effect, with the background torsion acting as the external magnetic field. We expect a generalization of these effects, in the case of our extended non-linear Dirac equation, to be relevant for the particle physics of the early Universe or inside ultra-compact astrophysical objects.

\section{Discussion and conclusion} \label{secV}

In this work we have studied the Einstein-Cartan-Dirac-Maxwell model implementing the $U(1)$-symmetry breaking and discussed its cosmological applications. The theoretical foundations of this model rely on fermionic (spinors) and bosonic (vector) fields minimally coupled to torsion of a Riemann-Cartan spacetime geometry. In this framework one is led to the Cartan equations relating the torsion tensor to the fundamental matter fields via the total matter spin tensor. Substituting torsion as a function of the matter field variables one obtains generalized Einstein-like, Dirac-like and Maxwell-like equations. This induces non-linear Dirac and electromagnetic dynamics with self-interactions (fermion-fermion, and boson-boson) and non-minimal fermion-boson interactions, and the resulting energy-momentum contributions for the gravitational equations.

Regarding cosmology, the ECDM model presented here gives rise to generalized Friedman dynamics coupled to bosonic and fermionic fields. The model is simplified if one takes an effective fluid description without needing to solve for the (generalized) Hehl-Datta-Dirac equation on a FLRW background. The resulting model predicts non-singular cosmologies with a bounce, similarly as in the original ECSK theory. In the $U(1)$-broken phase and neglecting bosonic self-interactions, there is an effective fluid component with energy density scaling as $\rho^{\rm corr}\sim \rho^{s}+\rho^{s-A}$, where $\rho^{s}\sim -\kappa^{2}\breve{s}^{2}\sim n^{2}\sim a^{-6}$ is a (negative) contribution from the spin-spin self interaction, and $\rho^{s-A}\sim\kappa^{2}\breve{s}^{2}f(A)$ comes from the non-minimal interactions (induced by torsion) between fermionic and bosonic fields. The latter can also introduce a negative contribution to the energy density depending on the $f(A)$ contribution (and therefore on the evolution of the bosonic 4-vector $A(a)$). We considered two different ansatze for the four-potential, namely $A_{\mu}=(\phi,0,0,0)$ and $A_{\mu}=(0,\vec{A})$. A typical example is $f(A)\sim \lambda\kappa^{2}A^{2}$, in the approximation where torsion is exclusively due to the spin tensor of fermions (Cases III and IV), although $\lambda^{2}\kappa^{4}A^{4}$ terms can also be present in the case where the bosonic spin tensor also contributes to torsion (Cases I and II). In all cases, we get a non-singular early Universe description in terms of a minimum value for the scale factor at which $H(a)=0$ for all possible spatial curvature values $k=-1,0,1$, due to the (negative) contribution from the spin-spin interaction. Moreover, these solutions show an accelerated expansion period after the bounce until $H(a)$ reaches a maximum value, followed by a decelerated expansion.

Regarding the effects of the non-minimal interactions induced by torsion, in the variation of Case IV we get $\rho^{s-A}\simeq {\rm constant} <0$, which has no significant effect on the early dynamics, but it can give rise to a halt of the decelerated expansion period at some future value of the scale factor, that is, $H(a_{max})=0$. In fact, most cases manifest this late-time behaviour for non-negligible values of $\lambda$. By considering the two branches of the family of solutions for the Hubble parameter, $H(a)=\pm\sqrt{\kappa^{2}/3(\rho+\rho^{\rm corr})-k/a^{2}}$, and the physical requirement of matter obeying the weak or dominant energy conditions, one is naturally led to interpret such future behaviour as a bounce (continuously) bridging a decelerated expansion phase to an accelerated contraction phase. Then, following this negative solution of the square root above, the accelerated contraction also reaches a maximum (absolute) value when the Hubble parameter reaches a (negative) minimum and  the contraction progresses in a decelerated manner until it reaches another minimum of the scale factor. At that instant, again the Hubble parameter $H(a)$ vanishes and the solution transits from the negative root to the positive root branch, in accordance with the physical energy (weak) conditions. This is another bounce, linking a decelerated contraction phase to an accelerated expansion and the cycle repeats over and over (see Fig. \ref{fig:2}).
This cyclic behaviour depends on both the existence of the strong spin-spin (negative energy) effect and on the (negative) energy contribution from the non-minimal couplings, which only becomes relevant in the late-time decelerated expansion phase. The strength of such a term depends on the single free parameter of the model, $\lambda$. The cyclic Universes are more intuitive for models where $f(A)\sim a^{n}$ with $n\geqslant 6$ (but are not exclusive to these), as long as $\rho^{\rm corr}<0$.

It is pertinent to briefly comment on the existence of negative energy densities in the ECDM model. Firstly, in the Friedmann equations (\ref{eq:F1}), the energy density $\rho$ of the relativistic fluid in the radiation era (the quark-gluon-lepton plasma) is positive. It is the corrections induced by torsion i.e, the spin-spin (fermion-fermion) self-interaction and the fermion-boson non-minimal couplings, that can be negative. The spin-spin energy density is always negative (also present in the Weyssenhof fluid of ECSK cosmology), and the induced fermion-boson non-minimal coupling can also be negative in some cases. As mentioned earlier, the first one is responsible for preventing the initial cosmic singularity and decays very rapidly with $\sim a^{-6}$, while the second one can give rise to late-time effects and in particular to the occurrence of a future cosmic bounce, with the corresponding transition from an expansion phase into a contraction one. The effective energy density should not be considered as a sum of different fluid components (since no extra fields are assumed), but rather as a single relativistic fluid with spin, that contains certain torsion-induced interaction energies within its fermionic and bosonic fields. Although these interactions can have negative energy densities associated to it, the analysis of cosmological perturbations of the Einstein-Cartan-Dirac-Maxwell theory is beyond the scope of the paper and might be addressed in a future work. The analysis of cosmological perturbations of the Weyssenhof spin fluid in ECSK, which also has spin-spin negative energy densities induced by torsion, has already been carried out (see for example \cite{Palle:1998qf}).

One of the solutions found (Case IV) is particularly interesting as it is a non-singular cosmology with an early acceleration period followed by a decelerated expansion and finally by a late-time accelerated epoch (see Fig. \ref{fig:2}). In general, all these late-time effects seem surprising, since usually one takes the torsion effects on the metric to be significant at or above Cartan's density $~10^{24}$g cm$^{-3}$. Although this is true for the (axial-axial) four-fermion spin-spin  self-interaction effects induced by torsion, the effects due to the non-minimal couplings in the matter fields induced by torsion can be relevant for late-time cosmology. The emergence in the same solution of bouncing early-time behaviour, an early period of accelerated expansion, a deceleration phase, and a late-time period of acceleration, is a fantastic example of the richness of the cosmological dynamics of an extremely simple theory  as the Einstein-Cartan theory with matter fields minimally coupled to the RC spacetime geometry.

The ECDM model predicts a negative cosmological constant in the variation of Case IV with an energy density scale set by $\lambda\kappa^{4}\hbar^{2}n^{2}_{\rm ref}A_{\rm ref}^{2}$. Such a constant is responsible for a cyclic cosmological behaviour as described above. On the other hand, if one takes a semi-quantum approach in the quark-gluon plasma and consider the presence of quark condensates in vacuum as predicted by QCD, i.e., the non-vanishing vacuum expectation value of $\left< 0|\bar{\psi}\psi|0\right>$, then the model predicts the existence of an effective cosmological constant and a dynamical dark energy contribution. The first term comes from the spin-spin energy interaction of vacuum (of fermionic quark fields) which enters the ECSK equations, and the second term is due to the (non-minimal) interaction between this vacuum term and the bosonic fields, taken here as classical fields. These results extend those of standard ECSK theory \cite{Poplawski:2010jv}, by adding a dynamical dark energy term which depends on the four-potential, and which cannot be neglected during the $U(1)$-broken symmetry phase induced by torsion. As long as the minimal coupling between torsion and the bosonic four-potential takes place, the dynamical dark energy term will be there. In other words, in the regimes in which the $U(1)$-breaking term in the bosonic Lagrangian (\ref{newMaxLagra}) is non-negligible the four-potential will evolve with the scale factor as derived from the corresponding Maxwell-like equations, or from the generalized continuity equations.

It is pertinent to ask when does the torsion ceases to be important and becomes negligible. The answer depends on the case: for instance, in the usual ECSK theory with torsion coupled only to fermions one gets $T_{\alpha\beta\gamma}\sim \kappa^{2}s^{D}_{\alpha\beta\gamma}\sim a^{-3}$ and the metric torsion effects scale with $\sim a^{-6}$, leading to a torsion era in the very early Universe. Now, when torsion couples also to vector bosonic fields, but it is only sourced by fermion spin density, then the $U(1)$-symmetry breaking Lagrangian term in (\ref{newMaxLagra}) can in principle decrease until it becomes negligible or not (see Table \ref{tab}). In the most general case, when torsion not only couples to the bosonic sector but it is also a result of the contribution from the total spin density including the spin density of bosons, the situation is similar to the case where torsion is due to fermionic spin densities, but there are situations in which a non-vanishing constant torsion background is predicted (variation of Case I). This topic requires further research since it needs to be carefully addressed in a quantum field theory context within a RC spacetime and strong-gravity regime.

From a more theoretical point of view, one can discuss the validity of using a description of matter in terms of fundamental fields (fermionic/bosonic) in the context of homogeneous and isotropic cosmologies and in the Einstein-Cartan theory and its extensions. On one hand, it should be reminded that the Weyssenhof fluid description can only be compatible with the cosmological principle upon some appropriate macroscopic averaging. On the other hand, it follows from a careful analysis of the paradigm changes that are required to consistently interpret the gauge theories of gravity with non-Riemann geometries, that Cartan equations are more appropriately interpreted as valid in microscopic scales.

One should also mention that the energy-momentum tensor terms derived from the non-minimal couplings in the matter Lagrangian could give rise to an effective fluid description which introduces anisotropic stresses. This should affect the dynamics via the Raychaudhuri equation and/or the conservation equation. We did not take into account such effects in the present work since we used the  assumptions $\bar{A^{i}A_{j}}\sim A^{2}\delta^{i}_{j}$ and $\bar{\breve{s}^{i}\breve{s}_{j}}\sim\breve{s}^{2}\delta^{i}_{j}$ to simplify the analysis. Again, this is reminiscent of the studies of the Weyssenhof fluid, which is not fully compatible with the cosmological principle, but can still be considered in the context of FLRW models by invoking macroscopic averaging arguments \cite{Obukhov}. Similarly, by exploring this idea, our model calls for a more self-consistent cosmological approach, for instance within Bianchi spacetimes. Alternatively, if one maintains the FLRW models at the background level, the perturbations should incorporate the anisotropic stresses, which might be important for the generation of cosmological GWs induced by spin density fluctuations (with non-zero, time varying quadrupole moment) in the early universe. One should also expect the production of GWs from the transitions between the primordial phases: from the $U(1)$-broken phase to the $U(1)$-restored phase (in particular, if this symmetry breaking is spontaneously induced rather than explicit), and from the usual torsion-dominated phase to the radiation phase. These transitions can contribute to a stochastic GW background of cosmological origin, with possible imprints from the physics beyond the standard model.

To conclude, in our view there are good motivations to keep with the analysis of gravitational models where non-Riemannian geometries, fermionic spin densities, and phase transitions become important, which can be tested with astrophysical and cosmological GW observations in the near future.  Work along these lines is currently underway.

\section*{Acknowledgments}

FC is funded by the Funda\c{c}\~ao para a Ci\^encia e a Tecnologia (FCT, Portugal) doctoral grant No.PD/BD/128017/2016.
FSNL acknowledges support from the FCT Scientific Employment Stimulus contract with reference
CEECIND/04057/2017.
DRG is funded by the \emph{Atracci\'on de Talento Investigador} programme of the Comunidad de Madrid (Spain) No. 2018-T1/TIC-10431, and acknowledges further support from the Ministerio de Ciencia, Innovaci\'on y Universidades (Spain) project No. PID2019-108485GB-I00/AEI/10.13039/501100011033, the Spanish project No. FIS2017-84440-C2-1-P (MINECO/FEDER, EU), and the Edital 006/2018 PRONEX (FAPESQ-PB/CNPQ, Brazil) Grant No. 0015/2019. The authors also acknowledge funding from FCT Projects No. UID/FIS/04434/2020, No. CERN/FIS- PAR/0037/2019 and No. PTDC/FIS- OUT/29048/2017.
FC thanks the hospitality of the Department of Theoretical Physics and IPARCOS of the Complutense University of Madrid, where part of this work was carried out. This article is based upon work from COST Actions CA15117  and CA18108, supported by COST (European Cooperation in Science and Technology).




\begin{thebibliography}{10}

   \bibitem{Will:2014kxa}
  C.~M.~Will,
  Living Rev.\ Rel.\  {\bf 17}, 4 (2014).

\bibitem{SgrA2017PhRvL.118u1101H}
 A.~Hees {\it et al.}, Phys.\ Rev.\ Lett.\ {\bf 118},  211101 (2017).


\bibitem{Abbott:2016blz}
B.~P.~Abbott {\it et al.}, Phys.\ Rev.\ Lett.\  {\bf 116},  061102 (2016).

\bibitem{Abbott:2017nn}
B. P. Abbott {\it et al.}, Phys. Rev. Lett. {\bf 119},  161101 (2017).

\bibitem{TheLIGOScientific:2016src}
  B.~P.~Abbott {\it et al.}
  Phys.\ Rev.\ Lett.\  {\bf 116},   221101 (2016)
   Erratum: [Phys.\ Rev.\ Lett.\  {\bf 121} (2018)  129902].

\bibitem{Abbott:2017oio}
B.~P.~Abbott {\it et al.}, Phys. Rev. Lett. \textbf{119}, 141101 (2017).

  \bibitem{TheLIGOScientific:2017qsa}
  B.~P.~Abbott {\it et al.}
  Phys.\ Rev.\ Lett.\  {\bf 119}, 161101 (2017).


\bibitem{M87EventHorizon}
The EHT Collaboration {\it et al.}, ApJL\  {\bf 875}, 1 (2019).

\bibitem{Bull:2015stt}
  P.~Bull {\it et al.},
  Phys.\ Dark Univ.\  {\bf 12} (2016) 56.

  \bibitem{Ezquiaga:2017ekz}
  J.~M.~Ezquiaga and M.~Zumalac\'arregui,
  Phys.\ Rev.\ Lett.\  {\bf 119} (2017) 251304.

   \bibitem{DeFelice:2010aj}
A.~De~Felice and S.~Tsujikawa, Living Rev. Rel. \textbf{13}, 3 (2010).

\bibitem{Olmoreview}
G. J. Olmo, Int. J. Mod. Phys. D \textbf{20}, 413 (2011)

 \bibitem{CLreview}
S.~Capozziello and M.~De Laurentis, Phys.\ Rept.\  {\bf 509}, 167 (2011).

\bibitem{NOOreview}
S.~Nojiri, S.~D.~Odintsov and V.~K.~Oikonomou,  Phys.\ Rept.\  {\bf 692}, 1 (2017).

  \bibitem{Berti:2015itd}
  E.~Berti {\it et al.},
  Class.\ Quant.\ Grav.\  {\bf 32}, 243001 (2015).

  \bibitem{BeltranJimenez:2017doy}
  J.~Beltran Jimenez, L.~Heisenberg, G.~J.~Olmo and D.~Rubiera-Garcia,
  Phys.\ Rept.\  {\bf 727}, 1 (2018).

\bibitem{Harko:2014gwa}
  T.~Harko and F.~S.~N.~Lobo,
  Galaxies {\bf 2}, 410 (2014).

\bibitem{Capozziello:2015lza}
  S.~Capozziello, T.~Harko, T.~S.~Koivisto, F.~S.~N.~Lobo and G.~J.~Olmo,
  Universe {\bf 1}, 199 (2015).

\bibitem{BookHarkoLobo}
T.~Harko and F.~S.~N.~Lobo, {\it Extensions of $f(R)$ Gravity: Curvature-Matter Couplings and Hybrid Metric-Palatini Theory} (CUP, 2018).


\bibitem{Blagojevic:2013xpa}
M.~Blagojevi\'c and F.~W.~Hehl,
  \emph{Gauge Theories of Gravitation: A Reader with Commentaries} (Imperial College Press, 2012).

\bibitem{Blagojevic}
  M.~Blagojevi\'c,
  \emph{Gravitation and Gauge symmetries} (Institute of Physics Publishing, 2002).

\bibitem{Obukhov}
  V.~N.~Ponomarev, A.~O.~Barvinsky, Y.~N.~Obukhov,
  \emph{Gauge approach and quantization methods in gravity theory} (Nauka, 2017).

\bibitem{Linde:1978px}
  A.~D.~Linde,
  Rept.\ Prog.\ Phys.\  {\bf 42},  389 (1979).

\bibitem{Zakout:2006zj}
  I.~Zakout, C.~Greiner and J.~Schaffner-Bielich,
  Nucl.\ Phys.\ A {\bf 781}, 150 (2007).

  \bibitem{Grojean:2006bp}
  C.~Grojean and G.~Servant,
  Phys.\ Rev.\ D {\bf 75},  043507 (2007).

  \bibitem{Lello:2014yha}
  L.~Lello and D.~Boyanovsky,
  Phys.\ Rev.\ D {\bf 91}, 063502 (2015).


  \bibitem{Barriga:2000nk}
  J.~Barriga, E.~Gaztanaga, M.~G.~Santos and S.~Sarkar,
  Mon.\ Not.\ Roy.\ Astron.\ Soc.\  {\bf 324},  977 (2001).



\bibitem{Obukhov:2018bmf}
  Y.~N.~Obukhov,
  Int.\ J.\ Geom.\ Meth.\ Mod.\ Phys.\  {\bf 15},  1840005 (2018).


\bibitem{Poplawski:2011xf}
  N.~J.~Poplawski,
  Phys.\ Rev.\ D {\bf 83}, 084033 (2011).


\bibitem{Cabral:2019gzh}
  F.~Cabral, F.~S.~N.~Lobo and D.~Rubiera-Garcia,
  Eur.\ Phys.\ J.\ C {\bf 79}, 1023 (2019).

\bibitem{Obukhov:2006gea}
  Y.~N.~Obukhov,
  Int.\ J.\ Geom.\ Meth.\ Mod.\ Phys.\  {\bf 3}, 95 (2006).



\bibitem{Poplawski:2011jz}
  N.~J.~Poplawski,
  Phys.\ Rev.\ D {\bf 85}, 107502 (2012).

  \bibitem{Unger:2018oqo}
  G.~Unger and N.~Poplawski, Astrophys. J. \textbf{870},  78 (2019).

\bibitem{Kranas:2018jdc}
  D.~Kranas, C.~G.~Tsagas, J.~D.~Barrow and D.~Iosifidis, arXiv:1809.10064 [gr-qc].

\bibitem{Poplawski:2010kb}
  N.~J.~Poplawski,
  Phys.\ Lett.\ B {\bf 694}, 181 (2010);
   Erratum: [Phys.\ Lett.\ B {\bf 701}, 672 (2011)].



\bibitem{Razina:2010bj}
  O.~Razina, Y.~Myrzakulov, N.~Serikbayev, G.~Nugmanova and R.~Myrzakulov,
  Central Eur.\ J.\ Phys.\  {\bf 10}, 47 (2012).

\bibitem{Palle:2014goa}
  D.~Palle,
  J.\ Exp.\ Theor.\ Phys.\  {\bf 118}, 587 (2014).

\bibitem{Poplawski:2012qy}
  N.~J.~Poplawski,
  arXiv:1201.0316 [astro-ph.CO]

\bibitem{Vakili:2013fra}
  B.~Vakili and S.~Jalalzadeh,
  Phys.\ Lett.\ B {\bf 726}, 28 (2013).

\bibitem{Xue:2008qs}
  S.~S.~Xue,
  Phys.\ Lett.\ B {\bf 665}, 54 (2008).


\bibitem{Ivanov:2016xjm}
  A.~N.~Ivanov and M.~Wellenzohn,
  Astrophys.\ J.\  {\bf 829}, 47 (2016).

  \bibitem{Cai:2015emx}
  Y.~F.~Cai, S.~Capozziello, M.~De Laurentis and E.~N.~Saridakis,
  Rept.\ Prog.\ Phys.\  {\bf 79}, 106901 (2016).

  \bibitem{Chen:2010va}
  S.~H.~Chen, J.~B.~Dent, S.~Dutta and E.~N.~Saridakis,
  Phys.\ Rev.\ D {\bf 83},  023508 (2011).

  \bibitem{Li:2011wu}
  B.~Li, T.~P.~Sotiriou and J.~D.~Barrow,
  Phys.\ Rev.\ D {\bf 83}, 104017 (2011).

  \bibitem{Bamba:2012vg}
  K.~Bamba, R.~Myrzakulov, S.~Nojiri and S.~D.~Odintsov,
  Phys.\ Rev.\ D {\bf 85}, 104036 (2012).

  \bibitem{Rodrigues:2012qua}
  M.~E.~Rodrigues, M.~J.~S.~Houndjo, D.~Saez-Gomez and F.~Rahaman,
  Phys.\ Rev.\ D {\bf 86}, 104059 (2012).

    \bibitem{Harko:2014sja}
  T.~Harko, F.~S.~N.~Lobo, G.~Otalora and E.~N.~Saridakis,
  Phys.\ Rev.\ D {\bf 89}, 124036 (2014).

\bibitem{Harko:2014aja}
  T.~Harko, F.~S.~N.~Lobo, G.~Otalora and E.~N.~Saridakis,
  JCAP {\bf 1412}, 021 (2014).

\bibitem{Carloni:2015lsa}
  S.~Carloni, F.~S.~N.~Lobo, G.~Otalora and E.~N.~Saridakis,
  Phys.\ Rev.\ D {\bf 93}, 024034 (2016).

\bibitem{Saez-Gomez:2016wxb}
  D.~Saez-Gomez, C.~S.~Carvalho, F.~S.~N.~Lobo and I.~Tereno,
  Phys.\ Rev.\ D {\bf 94}, 024034 (2016).

\bibitem{Khanapurkar:2018gyo}
  S.~Khanapurkar, A.~Pradhan, V.~Dhruv and T.~P.~Singh,
Phys. Rev. D \textbf{98}, 104027 (2018).

\bibitem{Khanapurkar:2018jvx}
  S.~Khanapurkar, A.~Varma, N.~Mittal, N.~Gupta and T.~P.~Singh,
  Phys.\ Rev.\ D {\bf 98}, 064046 (2018).

\bibitem{Lucat:2015rla}
  S.~Lucat and T.~Prokopec,
  JCAP {\bf 1710}, 047 (2017).

\bibitem{Poplawski:2011wj}
  N.~J.~Poplawski,
  Gen.\ Rel.\ Grav.\  {\bf 44},  491 (2012).




\bibitem{Poplawski:2010jv}
  N.~J.~Poplawski, Annalen Phys.\  {\bf 523},  291 (2011).

\bibitem{Poplawski:2009su}
  N.~J.~Poplawski,
  Phys.\ Lett.\ B {\bf 690},  73 (2010);
   Erratum: [Phys.\ Lett.\ B {\bf 727}, 575 (2013)].
   
   \bibitem{Jimenez:2020dpn}
J.~B.~Jim\'enez and A.~Delhom,
[arXiv:2004.11357 [gr-qc]].


\bibitem{Hehl-Data}
 F.~W.~Hehl,
Journal of Mathematical Physics, {\bf 12}, 1334 (1971).

  \bibitem{Iihoshi:2007uz}
  M.~Iihoshi, S.~V.~Ketov and A.~Morishita,
  Prog.\ Theor.\ Phys.\  {\bf 118}, 475 (2007).

  \bibitem{Gron1}
 O.~Gron and S.~Johannesen,
   Eur.\ Phys.\ J.\ Plus {\bf 126}, 28 (2011).

\bibitem{Gron2}
 O.~Gron and S.~Johannesen,  Eur.\ Phys.\ J.\ Plus {\bf 126}, 29 (2011).

\bibitem{Gron3}
 O.~Gron and S.~Johannesen,
  Eur.\ Phys.\ J.\ Plus {\bf 126}, 30 (2011).

\bibitem{Palle:1998qf}
 D.~Palle,
  Nuovo\ Cim.\ B. {\bf 114}, 853 (1999).

\end{thebibliography}
\end{document}